\documentclass[journal]{IEEEtran}
\usepackage[dvips]{graphicx,color}
\usepackage{latexsym}
\usepackage{amssymb}
\usepackage{amsmath, bm}
\usepackage{array}

\begin{document}
%
%
%
%
\newcommand{\qed}{\hfill$\square$}
\newcommand{\suchthat}{\mbox{~s.t.~}}
\newcommand{\markov}{\leftrightarrow}
%
%
\newenvironment{pRoof}{%
 \noindent{\em Proof.\ }}{%
 \hspace*{\fill}\qed \\
 \vspace{2ex}}


\newcommand{\ket}[1]{| #1 \rangle}
\newcommand{\bra}[1]{\langle #1 |}
\newcommand{\bol}[1]{\mathbf{#1}}
\newcommand{\rom}[1]{\mathrm{#1}}
\newcommand{\san}[1]{\mathsf{#1}}
\newcommand{\mymid}{:~}
\newcommand{\argmax}{\mathop{\rm argmax}\limits}
\newcommand{\argmin}{\mathop{\rm argmin}\limits}
%
%
%
%
\newcommand{\bc}{\begin{center}}  %
\newcommand{\ec}{\end{center}}
\newcommand{\befi}{\begin{figure}[h]}  %
\newcommand{\enfi}{\end{figure}}
\newcommand{\bsb}{\begin{shadebox}\begin{center}}   %
\newcommand{\esb}{\end{center}\end{shadebox}}
\newcommand{\bs}{\begin{screen}}     %
\newcommand{\es}{\end{screen}}
\newcommand{\bib}{\begin{itembox}}   %
\newcommand{\eib}{\end{itembox}}
\newcommand{\bit}{\begin{itemize}}   %
\newcommand{\eit}{\end{itemize}}
\newcommand{\defeq}{\stackrel{\triangle}{=}}
\newcommand{\Qed}{\hbox{\rule[-2pt]{3pt}{6pt}}}
\newcommand{\beq}{\begin{equation}}
\newcommand{\eeq}{\end{equation}}
\newcommand{\beqa}{\begin{eqnarray}}
\newcommand{\eeqa}{\end{eqnarray}}
\newcommand{\beqno}{\begin{eqnarray*}}
\newcommand{\eeqno}{\end{eqnarray*}}
\newcommand{\ba}{\begin{array}}
\newcommand{\ea}{\end{array}}
\newcommand{\vc}[1]{\mbox{\boldmath $#1$}}
\newcommand{\lvc}[1]{\mbox{\scriptsize \boldmath $#1$}}
\newcommand{\svc}[1]{\mbox{\scriptsize\boldmath $#1$}}

\newcommand{\wh}{\widehat}
\newcommand{\wt}{\widetilde}
\newcommand{\ts}{\textstyle}
\newcommand{\ds}{\displaystyle}
\newcommand{\scs}{\scriptstyle}
\newcommand{\vep}{\varepsilon}
\newcommand{\rhp}{\rightharpoonup}
\newcommand{\cl}{\circ\!\!\!\!\!-}
\newcommand{\bcs}{\dot{\,}.\dot{\,}}
\newcommand{\eqv}{\Leftrightarrow}
\newcommand{\leqv}{\Longleftrightarrow}
\newtheorem{co}{Corollary} 
\newtheorem{lm}{Lemma} 
\newtheorem{Ex}{Example} 
\newtheorem{Th}{Theorem}
\newtheorem{df}{Definition} 
\newtheorem{pr}{Property} 
\newtheorem{pro}{Proposition} 
\newtheorem{rem}{Remark} 

\newcommand{\lcv}{convex } 

\newcommand{\hugel}{{\arraycolsep 0mm
                    \left\{\ba{l}{\,}\\{\,}\ea\right.\!\!}}
\newcommand{\Hugel}{{\arraycolsep 0mm
                    \left\{\ba{l}{\,}\\{\,}\\{\,}\ea\right.\!\!}}
\newcommand{\HUgel}{{\arraycolsep 0mm
                    \left\{\ba{l}{\,}\\{\,}\\{\,}\vspace{-1mm}
                    \\{\,}\ea\right.\!\!}}
\newcommand{\huger}{{\arraycolsep 0mm
                    \left.\ba{l}{\,}\\{\,}\ea\!\!\right\}}}
\newcommand{\Huger}{{\arraycolsep 0mm
                    \left.\ba{l}{\,}\\{\,}\\{\,}\ea\!\!\right\}}}
\newcommand{\HUger}{{\arraycolsep 0mm
                    \left.\ba{l}{\,}\\{\,}\\{\,}\vspace{-1mm}
                    \\{\,}\ea\!\!\right\}}}

\newcommand{\hugebl}{{\arraycolsep 0mm
                    \left[\ba{l}{\,}\\{\,}\ea\right.\!\!}}
\newcommand{\Hugebl}{{\arraycolsep 0mm
                    \left[\ba{l}{\,}\\{\,}\\{\,}\ea\right.\!\!}}
\newcommand{\HUgebl}{{\arraycolsep 0mm
                    \left[\ba{l}{\,}\\{\,}\\{\,}\vspace{-1mm}
                    \\{\,}\ea\right.\!\!}}
\newcommand{\hugebr}{{\arraycolsep 0mm
                    \left.\ba{l}{\,}\\{\,}\ea\!\!\right]}}
\newcommand{\Hugebr}{{\arraycolsep 0mm
                    \left.\ba{l}{\,}\\{\,}\\{\,}\ea\!\!\right]}}
\newcommand{\HUgebr}{{\arraycolsep 0mm
                    \left.\ba{l}{\,}\\{\,}\\{\,}\vspace{-1mm}
                    \\{\,}\ea\!\!\right]}}

\newcommand{\hugecl}{{\arraycolsep 0mm
                    \left(\ba{l}{\,}\\{\,}\ea\right.\!\!}}
\newcommand{\Hugecl}{{\arraycolsep 0mm
                    \left(\ba{l}{\,}\\{\,}\\{\,}\ea\right.\!\!}}
\newcommand{\hugecr}{{\arraycolsep 0mm
                    \left.\ba{l}{\,}\\{\,}\ea\!\!\right)}}
\newcommand{\Hugecr}{{\arraycolsep 0mm
                    \left.\ba{l}{\,}\\{\,}\\{\,}\ea\!\!\right)}}

\newcommand{\hugepl}{{\arraycolsep 0mm
                    \left|\ba{l}{\,}\\{\,}\ea\right.\!\!}}
\newcommand{\Hugepl}{{\arraycolsep 0mm
                    \left|\ba{l}{\,}\\{\,}\\{\,}\ea\right.\!\!}}
\newcommand{\hugepr}{{\arraycolsep 0mm
                    \left.\ba{l}{\,}\\{\,}\ea\!\!\right|}}
\newcommand{\Hugepr}{{\arraycolsep 0mm
                    \left.\ba{l}{\,}\\{\,}\\{\,}\ea\!\!\right|}}

\newcommand{\MEq}[1]{\stackrel{
{\rm (#1)}}{=}}

\newcommand{\MLeq}[1]{\stackrel{
{\rm (#1)}}{\leq}}

\newcommand{\ML}[1]{\stackrel{
{\rm (#1)}}{<}}

\newcommand{\MGeq}[1]{\stackrel{
{\rm (#1)}}{\geq}}

\newcommand{\MG}[1]{\stackrel{
{\rm (#1)}}{>}}

\newcommand{\MPreq}[1]{\stackrel{
{\rm (#1)}}{\preceq}}

\newcommand{\MSueq}[1]{\stackrel{
{\rm (#1)}}{\succeq}}

\newcommand{\Ch}{{\Gamma}}
\newcommand{\Rw}{{W}}

\newcommand{\Cd}{{\cal R}_{\rm d}(\Ch)}
\newcommand{\CdB}{{\cal R}_{\rm d}^{\prime}(\Ch)}
\newcommand{\CdBB}{{\cal R}_{\rm d}^{\prime\prime}(\Ch)}

\newcommand{\Cdi}{{\cal R}_{\rm d}^{\rm (in)}(\Ch)}
\newcommand{\Cdo}{{\cal R}_{\rm d}^{\rm (out)}(\Ch)}

\newcommand{\tCdi}{\tilde{\cal R}_{\rm d}^{\rm (in)}(\Ch)}
\newcommand{\tCdo}{\tilde{\cal R}_{\rm d}^{\rm (out)}(\Ch)}
\newcommand{\hCdo}{  \hat{\cal R}_{\rm d}^{\rm (out)}(\Ch)}

\newcommand{\Cs}{{\cal R}_{\rm s}(\Ch)}
\newcommand{\CsB}{{\cal R}_{\rm s}^{\prime}(\Ch)}
\newcommand{\CsBB}{{\cal R}_{\rm s}^{\prime\prime}(\Ch)}

\newcommand{\Csi}{{\cal R}_{\rm s}^{\rm (in)}(\Ch)}
\newcommand{\Cso}{{\cal R}_{\rm s}^{\rm (out)}(\Ch)}
\newcommand{\tCsi}{\tilde{\cal R}_{\rm s}^{\rm (in)}(\Ch)}
\newcommand{\tCso}{\tilde{\cal R}_{\rm s}^{\rm (out)}(\Ch)}
\newcommand{\cCsi}{\check{\cal R}_{\rm s}^{\rm (in)}(\Ch)}
\newcommand{\Cds}{{\cal C}_{\rm ds}(\Ch)}
\newcommand{\Cdsi}{{\cal C}_{\rm ds}^{\rm (in)}(\Ch)}
\newcommand{\Cdso}{{\cal C}_{\rm ds}^{\rm (out)}(\Ch)}
\newcommand{\tCdsi}{\tilde{\cal C}_{\rm ds}^{\rm (in)}(\Ch)}
\newcommand{\tCdso}{\tilde{\cal C}_{\rm ds}^{\rm (out)}(\Ch)}
\newcommand{\hCdso}{\hat{\cal C}_{\rm ds}^{\rm (out)}(\Ch)}
\newcommand{\Css}{{\cal C}_{\rm ss}(\Ch)}
\newcommand{\Cssi}{{\cal C}_{\rm ss}^{\rm (in)}(\Ch)}
\newcommand{\Csso}{{\cal C}_{\rm ss}^{\rm (out)}(\Ch)}
\newcommand{\tCssi}{\tilde{\cal C}_{\rm ss}^{\rm (in)}(\Ch)}
\newcommand{\tCsso}{\tilde{\cal C}_{\rm ss}^{\rm (out)}(\Ch)}
\newcommand{\Cde}{{\cal R}_{\rm d1e}(\Ch)}
\newcommand{\Cdei}{{\cal R}_{\rm d1e}^{\rm (in)}(\Ch)}
\newcommand{\Cdeo}{{\cal R}_{\rm d1e}^{\rm (out)}(\Ch)}
\newcommand{\tCdei}{\tilde{\cal R}_{\rm d1e}^{\rm (in)}(\Ch)}
\newcommand{\tCdeo}{\tilde{\cal R}_{\rm d1e}^{\rm (out)}(\Ch)}
\newcommand{\hCdeo}{  \hat{\cal R}_{\rm d1e}^{\rm (out)}(\Ch)} 
\newcommand{\Cse}{{\cal R}_{\rm s1e}(\Ch)}
\newcommand{\Csei}{{\cal R}_{\rm s1e}^{\rm (in)}(\Ch)}
\newcommand{\Cseo}{{\cal R}_{\rm s1e}^{\rm (out)}(\Ch)}
\newcommand{\tCsei}{\tilde{\cal R}_{\rm s1e}^{\rm (in)}(\Ch)}
\newcommand{\tCseo}{\tilde{\cal R}_{\rm s1e}^{\rm (out)}(\Ch)}

\newcommand{\Capa}{C}

\newcommand{\ZeTa}{\zeta(S;Y,Z|U)}
\newcommand{\ZeTaI}{\zeta(S_i;Y_i,Z_i|U_i)}

\newcommand{\SP}{\mbox{{\scriptsize sp}}}
\newcommand{\mSP}{\mbox{{\scriptsize sp}}}
\newcommand{\CEreg}{\irBr{rate} }
\newcommand{\CEregB}{rate\MarkOh{-equivocation }}

\newcommand{\Cls}{class NL}
\newcommand{\vSpa}{\vspace{0.3mm}}
\newcommand{\Prmt}{\zeta}
\newcommand{\pj}{\omega_n}

\newfont{\bg}{cmr10 scaled \magstep4}
\newcommand{\bigzerol}{\smash{\hbox{\bg 0}}}
\newcommand{\bigzerou}{\smash{\lower1.7ex\hbox{\bg 0}}}
\newcommand{\nbn}{\frac{1}{n}}
\newcommand{\ra}{\rightarrow}
\newcommand{\la}{\leftarrow}
\newcommand{\ldo}{\ldots}
\newcommand{\ep}{\epsilon }
\newcommand{\typi}{A_{\epsilon }^{n}}
\newcommand{\bx}{\hspace*{\fill}$\Box$}
\newcommand{\pa}{\vert}
\newcommand{\ignore}[1]{}
%
%

%
%

%
%
%

\newenvironment{jenumerate}
	{\begin{enumerate}\itemsep=-0.25em \parindent=1zw}{\end{enumerate}}
\newenvironment{jdescription}
	{\begin{description}\itemsep=-0.25em \parindent=1zw}{\end{description}}
\newenvironment{jitemize}
	{\begin{itemize}\itemsep=-0.25em \parindent=1zw}{\end{itemize}}
\renewcommand{\labelitemii}{$\cdot$}

\newcommand{\iro}[2]{{\color[named]{#1}#2\normalcolor}}
\newcommand{\irr}[1]{{\color[named]{Black}#1\normalcolor}}

\newcommand{\irg}[1]{{\color[named]{Green}#1\normalcolor}}

\newcommand{\irb}[1]{{\color[named]{Black}#1\normalcolor}}

\newcommand{\irBl}[1]{{\color[named]{Black}#1\normalcolor}}
\newcommand{\irWh}[1]{{\color[named]{White}#1\normalcolor}}

\newcommand{\irY}[1]{{\color[named]{Yellow}#1\normalcolor}}
\newcommand{\irO}[1]{{\color[named]{Orange}#1\normalcolor}}
\newcommand{\irBr}[1]{{\color[named]{Purple}#1\normalcolor}}
\newcommand{\IrBr}[1]{{\color[named]{Purple}#1\normalcolor}}
\newcommand{\irBw}[1]{{\color[named]{Brown}#1\normalcolor}}
\newcommand{\irPk}[1]{{\color[named]{Magenta}#1\normalcolor}}
\newcommand{\irCb}[1]{{\color[named]{CadetBlue}#1\normalcolor}}

\newcommand{\irMho}[1]{{\color[named]{Mahogany}#1\normalcolor}}
\newcommand{\irOlg}[1]{{\color[named]{Black}#1\normalcolor}}

\newcommand{\irBg}[1]{{\color[named]{BlueGreen}#1\normalcolor}}
\newcommand{\irCy}[1]{{\color[named]{Cyan}#1\normalcolor}}
\newcommand{\irRyp }[1]{{\color[named]{RoyalPurple}#1\normalcolor}}

\newcommand{\irAqm}[1]{{\color[named]{Aquamarine}#1\normalcolor}}
\newcommand{\irRyb}[1]{{\color[named]{RoyalBule}#1\normalcolor}}
\newcommand{\irNvb}[1]{{\color[named]{NavyBlue}#1\normalcolor}}
\newcommand{\irSkb}[1]{{\color[named]{SkyBlue}#1\normalcolor}}
\newcommand{\irTeb}[1]{{\color[named]{TeaBlue}#1\normalcolor}}
\newcommand{\irSep}[1]{{\color[named]{Sepia}#1\normalcolor}}
\newcommand{\irReo}[1]{{\color[named]{RedOrange}#1\normalcolor}}
\newcommand{\irRur}[1]{{\color[named]{RubineRed}#1\normalcolor}}
\newcommand{\irSa }[1]{{\color[named]{Salmon}#1\normalcolor}}
\newcommand{\irAp}[1]{{\color[named]{Apricot}#1\normalcolor}}

%
\newenvironment{indention}[1]{\par
\addtolength{\leftskip}{#1}\begingroup}{\endgroup\par}
%
\newcommand{\namelistlabel}[1]{\mbox{#1}\hfill} 
\newenvironment{namelist}[1]{%
\begin{list}{}
{\let\makelabel\namelistlabel
\settowidth{\labelwidth}{#1}
\setlength{\leftmargin}{1.1\labelwidth}}
}{%
\end{list}}
%
%
\newcommand{\bfig}{\begin{figure}[t]}
\newcommand{\efig}{\end{figure}}
\setcounter{page}{1}

\newtheorem{theorem}{Theorem}
\newcommand{\Ep}{\mbox{\rm e}}

\newcommand{\Exp}{\exp
}
\newcommand{\idenc}{{\varphi}_n}
\newcommand{\resenc}{
{\varphi}_n}
\newcommand{\ID}{\mbox{\scriptsize ID}}
\newcommand{\TR}{\mbox{\scriptsize TR}}
\newcommand{\Av}{\mbox{\sf E}}

\newcommand{\Vl}{|}
\newcommand{\Ag}{(R,P_{X^n}|W^n)}
\newcommand{\Agv}[1]{({#1},P_{X^n}|W^n)}
\newcommand{\Avw}[1]{({#1}|W^n)}

\newcommand{\Jd}{X^nY^n}
\newcommand{\IdR}{r_n}

\newcommand{\Index}{{n,i}}

\newcommand{\cid}{C_{\mbox{\scriptsize ID}}}
\newcommand{\cida}{C_{\mbox{{\scriptsize ID,a}}}}
\newcommand{\rmOH}{\empty
}

\newcommand{\NoizeVar}{\sigma^2}

\newcommand{\GN}{
\frac{{\rm e}^{-\frac{(y-x)^2}{2\NoizeVar}}}
{\sqrt{2\pi {\NoizeVar}}}}

\arraycolsep 0.5mm
\date{}


\newcommand{\BiBArXiv}{
\bibitem{SingleStConvArXiv17}
Y. Oohama, ``Exponent function for stationary memoryless channels 
with input cost at rates above the capacity,'' 
{\it preprint;} 
available at available at https://arxiv.org/abs/1701.06545.
}

\newcommand{\GauArXiv}{in \cite{SingleStConvGauArXiv17}. }
\newcommand{\ArXiv}{in \cite{SingleStConvArXiv17}. }
\newcommand{\GorF}{found }
\newcommand{\GorFb}{is found }
\newcommand{\Comment}{}
%
%
%
\title{
The Reliability Function for the Additive White Gaussian Noise Channel 
at Rates above the Capacity
}
\author{%
Yasutada Oohama 
\thanks{
Y. Oohama is with 
Dept. of Communication Engineering and Informatics,
University of Electro-Communications,
1-5-1 Chofugaoka Chofu-shi, Tokyo 182-8585, Japan.
}%
\thanks{
}
}
\markboth{
}
{
}
\maketitle

\begin{abstract}
We consider the additive white Gaussian noise channels. 
We prove that the error probability of decoding tends to one 
exponentially for rates above the capacity and derive 
the optimal exponent function. We shall demonstrate 
that the information spectrum approach is quite 
useful for investigating this problem.
\end{abstract}


\section{Introduction
}

It is well known that discrete memoryless channels(DMCs) have a property 
that the error probability of decoding goes to one as the block length 
$n$ of transmitted codes tends to infinity at rates above the channel 
capacity. This property is called the strong converse property. In this 
paper we study the strong converse property for additive white Gaussian 
noise channels(AWGNs). Han \cite{Han98InfSpec} proved that we have the 
strong converse property for AWGNs. In \cite{Han98InfSpec}, Han 
introduced a new method to study several coding problems of information 
theory. This method is called the information spectrum method. Based on 
the information spectrum method, Han \cite{Han98InfSpec} gave a simple 
proof of the strong property for AWGNs. 

In this paper for AWGNs we study an asymptotic behavior for the correct 
probability of decoding to vanish at rates above the capacity. To our 
knowledge we have had no work on this subject. In the case of DMCs, 
Arimoto \cite{ari} proved that the correct probability of decoding 
vanishes exponentially at rates above the capacity and derive an 
explicit lower bound of this exponent. Dueck and K\"orner 
\cite{Dueck_Korner1979} determined the optimal exponent. The equivalence 
of the bound of Arimoto \cite{ari} to the optimal bound of Dueck 
and K\"orner \cite{Dueck_Korner1979} was proved by Oohama 
\cite{OohamaIeice15}. Derivations of the optimal exponent using 
information spectrum method was investigated by Nagaoka \cite{Nagaoka01}, 
Hayashi and Nagaoka \cite{HayashiNagaoka03}, and Oohama 
\cite{SingleStConvArXiv17}. 

In this paper we determine the optimal exponent function on the correct 
probability of decoding at rates above capacity for AWGNs. To obtain 
this result we use a lower bound of the exponent function derived by 
Oohama \cite{SingleStConvArXiv17}. This lower bound is not computable 
since it has an expression of the variational problem. We solve 
this problem by using its min-max structure, obtaining an 
explicit formula of this lower bound. This formula coincides with an 
upper bound obtained by another argument, thereby establishing the 
optimal exponent. Those arguments are quite simple and elementary.

\section{
The Capacity of the Additive White Gaussian Noise Channels
}

\newcommand{\ZapGaus}{
In this section we consider 
the case where $W$ is an additive white Gaussian noise(AWGN). 
We assume that ${\cal X}, {\cal Y}$ are real lines and 
$X\in {\cal X}$ and $Y \in {\cal Y}$ are continuous random variables.
AWGN is defined by
$$
Y=X+N, X \perp N, N\sim {\cal N}(0,\NoizeVar).  
$$
In this case $W=W(y|x)=q_N(y-x)$ is a 
probability density function(p.d.f.) given by
$$
W(y|x)=q_N(y-x)=\frac{1}{\sqrt{2\pi\NoizeVar}}{\rm e}^{-\frac{(y-x)^2}{\NoizeVar}}.  
$$ 
}

Let ${\cal X}$ and ${\cal Y}$ be real lines and 
let $X$ $\in {\cal X},$ and $Y\in {\cal Y}$ 
be real valued random variables. 
The set ${\cal X}$ corresponds to a channel input and 
the set ${\cal Y}$ corresponds  to a channel output. 
We write a conditional probability density function(p.d.f.) 
of $Y$ on given $X$ as 
$$
W=\left\{W({y}|{x})\right\}_{(x,y)
\in {\cal X} \times {\cal Y}}.
$$
Let $N$ be a Gussian random variable with mean 0 and variance 
${\NoizeVar}$. 
The additive Gaussian noise channel is defined 
by 
$$
Y=X+N, \quad X \perp N,
$$
where $X \perp N$ stands for that $X$ and $N$ are independent.
In this case $W=W(y|x)=p_N(y-x)$ is a 
probability density function(p.d.f.) given by
$$
W(y|x)=p_N(y-x)=\frac{1}{\sqrt{2\pi\NoizeVar}}{\rm e}^{-\frac{(y-x)^2}{\NoizeVar}}.  
$$ 
Let $X^n=(X_1,X_2,\cdots,X_n)$ be a random vector 
taking values in ${\cal X}^n$. We write an element of ${\cal X}^n$ as 
$x^n=x_{1}x_{2}$$\cdots x_{n}.$ Similar notations are adopted for other random variables. 
Let $Y^n \in {\cal Y}^n$ be a random variable 
obtained as the channel output by connecting 
$X^n$ to the input of channel.
The additive white Gaussian noise(AWGN) channel we study in this paper 
is defined by 
\beq
Y_t=X_t+N_t, X_t \perp N_t, \mbox{ for }t=1,2,\cdots,n,
\label{eqn:AWGN}
\eeq
where $N_t,$ $t=1,2\cdots,n $ are independent Gussian random variables 
having the same distribution as $N$. 
We write a conditional probability p.d.f. of $Y^n$ on given $X^n$ as 
$$
W^n=
\left\{W^n({y}^n|{x}^n)\right\}_{(x^n,y^n)
\in {\cal X}^n \times {\cal Y}^n}.
$$
By the definition (\ref{eqn:AWGN}) of the AWGN channel, we have 
\beqno
W^n(y^n|x^n)=\prod_{t=1}^n p_{N}(y_t-x_t).
\eeqno
The AWGN channel is specified with $\NoizeVar$.
Let $K_n$ be uniformly distributed random variables 
taking values in message sets ${\cal K}_n $. 
The random variable $K_n$ is a message sent to the receiver.
A sender transforms $K_n$ into a transmitted sequence 
$X^n$ using an encoder function and sends it to the receiver. 
In this paper we assume that the encoder function $\varphi^{(n)}$ 
is a deterministic encoder. In this case, $\varphi^{(n)}$ 
is a one-to-one mapping from ${\cal K}_n$ into ${\cal X}^n$. 
The decoding function 
at the receiver is denoted by ${\psi}^{(n)}$. 
This function is formally defined by
$
{\psi}^{(n)}: {\cal Y}^{n} \to {\cal K}_n.
$
Let $c: {\cal X} \to [0,\infty)$ be a cost function.
The average cost on output of $\varphi^{(n)}$ 
must not exceed $\Gamma$. This condition is given by
$\varphi^{(n)}({\empty K}_n) \in {\cal S}_{\Gamma}^{(n)}$, where 
\beqno
{\cal S}_{\Gamma}^{(n)} 
&\defeq & \biggl\{x^n\in {\cal X}^n: 
\frac{1}{n}\sum_{t=1}^n c(x_t) \leq \Gamma 
\biggr\}.
\eeqno
We consider the case where the input constrain $c(X)$ 
is a power constraint give by $c(x)=x^2$. 
The average error probabilities of decoding at 
the receiver is defined by 
\beqno
{\rm P}_{\rm e}^{(n)}
&=&{\rm P}_{\rm e}^{(n)}(\varphi^{(n)},\psi^{(n)}|{W})
\defeq \Pr\{ \psi^{(n)}(Y^n)\neq K_n \}
\\
&=&1-\Pr\{\psi^{(n)}(Y^n)= K_n \}.
\eeqno
For $k\in {\cal K}_n$, set
$
{\cal D}(k)\defeq \{ y^n: \psi^{(n)}(y^n)=k \}.
$
The families of sets 
$\{ {\cal D}(k) \}_{k\in {\cal K}_n}$ 
is called the decoding regions. Using the decoding region, 
${\rm P}_{\rm e}^{(n)}$ can be written as
\beqno
& &{\rm P}_{\rm e}^{(n)}={\rm P}_{\rm e}^{(n)}(\varphi^{(n)},\psi^{(n)}{}|W) 
\\
&=&\frac{1}{|{\cal K}_n|} 
\sum_{k\in {\cal K}_n }
\int_{\scs y^n \notin {\cal D}(k)
    }
{\rm d}{y^n}
W^n\left(y^n\left| \varphi^{(n)}(k) \right.\right)
\\
&=&\frac{1}{|{\cal K}_n|} 
\sum_{k\in {\cal K}_n }
\int_{\scs y^n \notin {\cal D}(k)
    }
{\rm d}{y^n}
\prod_{t=1}^n W\left(y_t\left|x_t(k)\right.\right),
\eeqno
where $x_t(k)=[\varphi^{(n)}(k)]_t$, $t=1,2,\cdots,n$ 
are the $t$-th components of $x^n=x^n(k)$ $=\varphi^{(n)}(k)$
and $\pa {\cal K}_n \pa$ is a cardinality of the set ${\cal K}_n$. 
Set
\beqno
& &{\rm P}^{(n)}_{\rm c}
={\rm P}^{(n)}_{\rm c}(\varphi^{(n)},\psi^{(n)}|W)
\defeq 
1-{\rm P}^{(n)}_{\rm e}(\varphi^{(n)}, \psi^{(n)}|W)
\eeqno
The quantity ${\rm P}^{(n)}_{\rm c}$ is called the average correct 
probability of decoding. This quantity has the following form
\beqno
& &{\rm P}_{\rm c}^{(n)}
={\rm P}_{\rm c}^{(n)}(\varphi^{(n)},\psi^{(n)}|W)
\\
&=&\frac{1}{|{\cal K}_n|} 
\sum_{k\in {\cal K}_n }
\int_{\scs  y^n
        \in 
{\cal D}(k)
    }
{\rm d}{y^n}
W^n\left( y^n  \left| \varphi^{(n)}(k) \right.\right).
\eeqno
For given $\varepsilon$ $\in (0,1)$, 
$R$ is $\varepsilon$-{\it achievable} under $\Gamma$
if for any $\delta>0$, there exist a positive integer 
$n_0=n_0(\varepsilon,\delta)$ and a sequence of pairs 
$
\{(\varphi^{(n)},\psi^{(n)}): 
\varphi^{(n)}({\empty K}_n) \in {\cal S}_{\Gamma}^{(n)} \}_{n=1}^{\infty}
$ 
such that for any $n\geq n_0(\varepsilon,\delta)$, 
\beqa 
{\rm P}_{{\rm e}}^{(n)}
(\varphi^{(n)},\psi^{(n)} | W )
&\leq &\varepsilon, 
\quad
\nbn \log \pa {\cal K}_n \pa  \geq  R-\delta.
\eeqa
The supremum of all $\varepsilon$-achievable 
$R$ under $\Gamma$ is denoted by 
${C}_{\rm AWGN}(\varepsilon, \Gamma | \NoizeVar)$.
We set
$$
{C}_{\rm AWGN}(\Gamma|\NoizeVar)
\defeq \inf_{\varepsilon\in (0,1)}
C_{\rm AWGN}(\varepsilon,\Gamma|\NoizeVar),
$$
which is called the channel capacity. 
The maximum error probability of decoding 
is defined by as follows:
\beqno
{\rm P}_{{\rm e,\irOlg{m}}}^{(n)}&=&
   {\rm P}_{{\rm e,\irOlg{m}}}^{(n)}(\varphi^{(n)},\psi^{(n)}|W)
\\
& \defeq & \max_{k \in {\cal K}_n}
\Pr\{\psi^{(n)}(Y^n)\neq k|K_n=k\}.
\eeqno
Based on this quantity, we define the \irOlg{maximum capacity} 
as follows. For a given $\varepsilon \in (0,1)$, 
$\irb{R}$ is $\varepsilon$-{\it achievable} under 
$\Gamma$, if for any $\delta>0$, there exist a positive integer 
$n_0=n_0(\varepsilon,\delta)$ and a sequence of pairs 
$
\{(\varphi^{(n)},\psi^{(n)}): 
\varphi^{(n)}({\empty K}_n) \in {\cal S}_{\Gamma}^{(n)} \}_{n=1}^{\infty}
$ 
such that for any $n\geq n_0(\varepsilon,\delta)$, 
\beqa 
{\rm P}_{{\rm e},\irOlg{\rm m}}^{(n)}
(\varphi^{(n)},\psi^{(n)} |W)
&\leq &\varepsilon, 
\quad
\nbn \log \pa {\cal K}_n \pa 
\geq \irb{R}-\delta.
\eeqa
The supremum of all $\varepsilon$-achievable 
rates under $\Gamma$ is denoted by $C_{\irOlg{\rm m},{\rm AWGN}}
(\varepsilon,\Gamma| \NoizeVar)$. 
We set 
$$
{C}_{\irOlg{\rm m},\rm AWGN}(\Gamma | \NoizeVar)
=\inf_{\varepsilon\in (0,1)} 
C_{\irOlg{\rm m},\rm AWGN}(\varepsilon,\Gamma |\NoizeVar)
$$
which is called the \irOlg{maximum} capacity of the AWGN.
Let ${\cal G}_{d}$ be a set of all $d$ dimensional 
Gaussian distributions. Set
\beq
C(\Gamma |\NoizeVar)=\max_{\scs p_X \in {\cal G}_1:
          \atop{\scs
           {\rm E}_{p_X}X^2 \leq \Gamma 
          }
     }I(p_X,W)=\frac{1}{2}\log\left(1+\frac{\Gamma}{\NoizeVar}\right). 
\eeq
where $I(p_X,W)$ stands for a mutual information 
between $X$ and $Y$ when input distribution of $X$ is $p_X$.
The following is a well known result.
\begin{Th}
\label{th:ddirect} 
{\rm For any  AWGN, we have
$$
{C}_{\rm m, AWGN}(\Gamma| \NoizeVar)={C}_{\rm AWGN}(\Gamma|\NoizeVar)
={C}(\Gamma| \NoizeVar).
$$
}
\end{Th}

Han \cite{Han98InfSpec} established the strong converse theorem for AWGNs.
His result is as follows. 

\begin{Th}[Han \cite{Han98InfSpec}] 
If $R>C(\Gamma | \NoizeVar)$, we have 
$$
\lim_{n\to\infty}{\rm P}_{\rm e}^{(n)}
(\varphi^{(n)}, \psi^{(n)}|{W})=1
$$ 
for any 
$\{(\varphi^{(n)},\psi^{(n)}): 
\varphi^{(n)}({\empty K}_n) \in {\cal S}_{\Gamma}^{(n)} \}_{n=1}^{\infty}
$ 
satisfying 
$$
\frac{1}{n}\liminf_{n\to \infty}M_n \geq R.
$$ 
\end{Th}

The following corollary immediately follows from this theorem. 

\begin{co}
For each fixed $\varepsilon$ $ \in (0,1)$ and any AWGN specified with 
$\NoizeVar$, we have 
$$
{C}_{\rm m, AWGN}(\varepsilon, \Gamma | {\NoizeVar})
={C}_{\rm AWGN}(\varepsilon,\Gamma | {\NoizeVar})={C}(\Gamma|\NoizeVar).
$$
\end{co}

To examine an asymptotic behavior of 
${\rm P}^{(n)}_{\rm c}(\varphi^{(n)}, \psi^{(n)}|W)$ for large $n$ 
at $R>C(\Gamma| {\NoizeVar})$, we define the following quantities:
\beqno
& & 
G^{(n)}(R,\Gamma| {\NoizeVar})
\\
&&\defeq
\min_{\scs 
(\varphi^{(n)},\psi^{(n)}):
    \atop{\scs 
         \varphi^{(n)}({\empty K}_n) \in {\cal S}_{\Gamma}^{(n)},
          \atop{\scs
               (1/n)\log M_n \geq R 
               }    
         }
    }
\hspace*{-2mm}
\left(-\frac{1}{n}\right)
\log {\rm P}_{\rm c}^{(n)}(\varphi^{(n)},\psi^{(n)}|{W}),
\\
& &
{G}^{*}(R,\Gamma| {\NoizeVar}) \defeq \lim_{n \to \infty} G^{(n)}(R,\Gamma| {\NoizeVar}).
\eeqno
On the above exponent functions, we have the 
following property. 
\begin{pr}\label{pr:pr0}{$\quad$
\begin{itemize}
\item[a)] By definition we have that for each fixed $n\geq 1$, $G^{(n)}(R,\Gamma| {\NoizeVar})$ 
is a monotone increasing function of $R\geq0$ and satisfies $G^{(n)}(R,\Gamma| {\NoizeVar}) \leq R$. 

\item[b)] The sequence $\{G^{(n)}(R,\Gamma| {\NoizeVar})$ $\}_{n\geq 1}$ of exponent functions  
satisfies the following subadditivity property:
\beqa
& &G^{(n+m)}(R,\Gamma|{\NoizeVar}) 
\nonumber\\
&\leq& \frac{nG^{(n)}(R,\Gamma| {\NoizeVar})+mG^{(m)}(R,\Gamma| {\NoizeVar})}{n+m},
\label{eqn:Zxcxx}
\eeqa
from which we have that ${G}^{*}(R,\Gamma| {\NoizeVar})$ exists and is equal to 
$\inf_{n\geq 1}G^{(n)}(R,\Gamma| {\NoizeVar})$. 

\item[c)]
For fixed $R>0$, the function ${G}^{*}(R,\Gamma| {\NoizeVar})$ is a monotone 
decreasing function 
of $\Gamma$. 
For fixed $\Gamma>\Gamma_0=\min_{x\in{\cal X}}c(x) $, the function 
${G}^{*}(R,\Gamma| {\NoizeVar})$ a monotone 
increasing function of $R$ and satisfies 
\beq
G^{*}(R,\Gamma|{\NoizeVar})\leq R.
\label{eqn:Zxc}
\eeq
\item[d)] The function ${G}^{*}(R,\Gamma| {\NoizeVar})$ is a convex function of 
$(R,\Gamma)$. 
\end{itemize}}
\end{pr}

Proof of this property 
is found in \cite{SingleStConvArXiv17}.
\newcommand{\ApdaAAAA}{
\subsection{
General Properties on ${G}^{*}(R,\Gamma| {\NoizeVar})$
} 
\label{sub:ApdaAAAA}

In this appendix we prove Property \ref{pr:pr0} describing general 
properties on ${G}^{*}(R,\Gamma| {\NoizeVar})$. 

{\it Proof of Property \ref{pr:pr0}:} 
By definition it is obvious that for fixed $\Gamma>0$, ${G}^{(n)}(R,\Gamma|{\NoizeVar})$ 
is a monotone increasing function of $R>0$ and that 
for fixed $R>0$, ${G}^{(n)}(R,\Gamma | {\NoizeVar})$ 
is a monotone increasing function of $\Gamma>0$. 
We prove the part b). By time sharing we have that 
\beqa
& &G^{(n+m)}\left(\left.
\frac{n R + m R^{\prime}}{n+m},
\frac{n \Gamma + m \Gamma^{\prime}}{n+m}
\right| {\NoizeVar}\right) 
\nonumber\\
&\leq& \frac{nG^{(n)}(R,\Gamma| {\NoizeVar})+mG^{(m)}(R^{\prime},\Gamma^{\prime}| {\NoizeVar})}{n+m}.
\label{eqn:Sdd} 
\eeqa
The part b) follows by letting $R=R^{\prime}$ and $\Gamma=\Gamma^{\prime}$ in (\ref{eqn:Sdd}). 
We next prove the part c). 
By definition it is obvious that for fixed $\Gamma>0$, ${G}^{*}(R,\Gamma | {\NoizeVar})$ 
is a monotone decreasing function of $R>0$ and that for fixed $R>0$, 
${G}^{*}(R,\Gamma | {\NoizeVar})$ is a monotone increasing function of $\Gamma>0$. 
It is obvious that the worst pair of $(\varphi^{(n)},\psi^{(n)})$ is 
that for $M_n=\lfloor {\rm e}^{nR}\rfloor$, the decoder $\psi^{(n)}$ 
always outputs a constant message $m_0 \in {\cal M}_n$.
In this case we have 
\beqno 
& &\lim_{n\to \infty }\left(-\frac{1}{n}\right)
\log {\rm P}_{\rm c}^{(n)}(\varphi^{(n)},\psi^{(n)}{}|{W})
\\
&=&\lim_{n\to \infty }\left(-\frac{1}{n}\right)\log {M_n}=R.
\eeqno
Hence  we have $(\ref{eqn:Zxc})$ in the part c). We finally prove the part d). 
Let $ \lfloor a  \rfloor $ be an integer part of $a$. Fix any $\alpha\in [0,1]$. 
Let $\bar{\alpha}=1-\alpha$. We choose $(n,m)$ so that 
$$
n=k_\alpha\defeq \lfloor k\alpha \rfloor,\: 
m=k_{\bar{\alpha}} \defeq  \lfloor k \bar{\alpha}\rfloor.
$$ 
For this choice of $n$ and $m$, we have
\beq
\left.
\ba{l}
\ds \left(1-\frac{1}{k}\right)\alpha \leq 
\frac{n}{n+m}\leq \frac{k}{k-1}\alpha
\vspace*{1mm}\\
\ds \left(1-\frac{1}{k}\right)\bar{\alpha} \leq 
\frac{m}{n+m}\leq \frac{k}{k-1}\bar{\alpha}\\
\ea
\right\}
\label{eqn:Sddaa} 
\eeq
Fix small positive $\tau$ arbitrary. Then, for any 
$$
k> \max\{ (\alpha R+\bar{\alpha}R^{\prime})/\tau,
          (\alpha \Gamma+\bar{\alpha}\Gamma^{\prime})/\tau\},
$$ 
we have the following chain of inequalities:
\beqa
& &G^{(k_\alpha+k_{\bar{\alpha}})}\left(\left.
\alpha R+\bar{\alpha}R^{\prime}-\tau, 
\alpha \Gamma +\bar{\alpha}\Gamma^{\prime}-\tau 
\right| {\NoizeVar}\right) 
\nonumber\\
&\MLeq{a}&
G^{(k_\alpha+k_{\bar{\alpha}})}\left(
\left(1-\frac{1}{k}\right)
\left(\alpha R+\bar{\alpha}R^{\prime}\right), \right.
\nonumber\\
& &\qquad \qquad \:\:\:
\left. \left. 
\left(1-\frac{1}{k}\right)
\left(\alpha \Gamma+\bar{\alpha}\Gamma^{\prime}\right)
\right| W\right) 
\nonumber\\
&\MLeq{b}&G^{(n+m)}\left(\left.
\frac{n R + m R^{\prime}}{n+m},
\frac{n \Gamma + m \Gamma^{\prime}}{n+m}
\right| {\NoizeVar}\right) 
\nonumber\\
&\MLeq{c}& \frac{nG^{(n)}(R,\Gamma| {\NoizeVar})+mG^{(m)}(R^{\prime},\Gamma^{\prime}| {\NoizeVar})}{n+m}
\nonumber\\
&\MLeq{d}&\left(\frac{k}{k-1}\right)\left[
\alpha G^{(k_\alpha)}(R,\Gamma| {\NoizeVar})
+\bar{\alpha}G^{(k_{\bar{\alpha}})}(R^{\prime},\Gamma^{\prime}| {\NoizeVar})\right].
\quad \:\label{eqn:Sddzzz}
\eeqa
Step (a) follows from the part a) and 
$$
k> \max\{ (\alpha R+\bar{\alpha}R^{\prime})/\tau,
          (\alpha \Gamma+\bar{\alpha}\Gamma^{\prime})/\tau\}.
$$ 
Step (b) follows from the part a).
Step (c) follows from (\ref{eqn:Sdd}).
Step (d) follows from (\ref{eqn:Sddaa}).
Letting $k\to\infty$ in (\ref{eqn:Sddzzz}), we have
\beqa
&  &G^{*}\left( \alpha R+\bar{\alpha} R^{\prime}-\tau, 
\alpha \Gamma+\bar{\alpha} \Gamma^{\prime}-\tau| {\NoizeVar} \right)
\nonumber\\
&\leq &\alpha G^*(R,\Gamma| {\NoizeVar}) 
+ \bar{\alpha} G^*(R^{\prime},\Gamma^{\prime}| {\NoizeVar}),
\label{eqn:adf}
\eeqa
where $\tau$ can be taken arbitrary small. 
We choose $R^{\prime}$, $\Gamma^{\prime}$, and $\alpha$, as 
\beq
\left.
\ba{l}
R^{\prime}=R+2\sqrt{\tau},\quad \Gamma^{\prime}=\Gamma +2\sqrt{\tau},
\\
\alpha =1-\sqrt{\tau}.
\ea
\right\}
\label{eqn:SddZ}
\eeq
For the above choice of $R^{\prime}$, $\Gamma^{\prime}$, and $\alpha$, we have 
\beq
\alpha R+\bar{\alpha} R^{\prime}=R+2\tau, \quad
\alpha \Gamma +\bar{\alpha} \Gamma^{\prime}=\Gamma +2\tau.
\label{eqn:SddZz}
\eeq
Then we have the following chain of inequalities:
\beqa
& &  G^{*}\left(R+\tau, \Gamma+\tau| {\NoizeVar} \right)
\nonumber\\
&\MEq{a}& G^{*}\left(\alpha R+\bar{\alpha} R^{\prime}-\tau,
              \alpha \Gamma +\bar{\alpha} \Gamma^{\prime}-\tau
| {\NoizeVar} \right)
\nonumber\\
&\MLeq{b}& \alpha G^*(R,\Gamma| {\NoizeVar})
+\bar{\alpha} G^*(R^{\prime},\Gamma^{\prime}| {\NoizeVar})
\nonumber\\
&\MLeq{c}& \alpha G^*(R,\Gamma| {\NoizeVar})+\bar{\alpha}R^{\prime}
\nonumber\\
&\MEq{d}&(1-\sqrt{\tau})G^*(R,\Gamma| {\NoizeVar})+\sqrt{\tau}
R+2\tau
\nonumber\\
&\leq &G^*(R,\Gamma| {\NoizeVar})+\sqrt{\tau}R+2\tau.
\label{eqn:adfZxxS}
\eeqa
Step (a) follows from (\ref{eqn:SddZz}). 
Step (b) follows from (\ref{eqn:adf}). 
Step (c) follows from (\ref{eqn:Zxc}). 
Step (d) follows from (\ref{eqn:SddZ}). 
For any positive $\tau$, we have the following chain of inequalities:
\beqa
&  & G^{*}\left(\alpha R+\bar{\alpha} 
R^{\prime},\alpha \Gamma+\bar{\alpha} \Gamma^{\prime}| {\NoizeVar} \right)
\nonumber\\
&=&G^{*}\left( \alpha R+\bar{\alpha} R^{\prime}
-\tau+\tau,\alpha \Gamma+\bar{\alpha} \Gamma^{\prime}-\tau+\tau | {\NoizeVar} \right)
\nonumber\\
&\MLeq{a} & G^*(\alpha R+\bar{\alpha} R^{\prime}-\tau,\alpha \Gamma+\bar{\alpha}
\Gamma^{\prime}-\tau | {\NoizeVar})
\nonumber\\
&   &   \qquad +\sqrt{\tau}(\alpha R+\bar{\alpha} R^{\prime}-\tau)+2\tau
\nonumber\\
&\MLeq{b} &\alpha G^*(R,\Gamma| {\NoizeVar}) + \bar{\alpha} G^*(R^{\prime},\Gamma^{\prime}| {\NoizeVar})
\nonumber\\
&   &   \qquad +\sqrt{\tau}(\alpha R+\bar{\alpha} R^{\prime})
+\tau(2-\sqrt{\tau}).
\label{eqn:adfZxxxS}
\eeqa
Step (a) follows from (\ref{eqn:adfZxxS}). 
Step (b) follows from (\ref{eqn:adf}). 
Since $\tau>0$ can be taken arbitrary small in (\ref{eqn:adfZxxxS}), we have  
\beqno
& & G^{*}\left( \alpha R+\bar{\alpha} R^{\prime},\alpha \Gamma+\bar{\alpha} \Gamma^{\prime}| {\NoizeVar} \right)
\\
&\leq &\alpha G^*(R,\Gamma| {\NoizeVar}) + \bar{\alpha} G^*(R^{\prime},\Gamma^{\prime}| {\NoizeVar}),
\eeqno
which implies the convexity of $G^*(R,\Gamma| {\NoizeVar})$ on $(R,\Gamma)$. 
\hfill\IEEEQED
}

\section{Main Results}
\label{sec:MainResult}

In this section we state our main results. We first 
present a result on an upper bound of $G^*(R,\Gamma |\NoizeVar)$. 
Define
\begin{align*}
& \overline{G}_{\rm DK}(R,\Gamma|{\NoizeVar}) \\
&\defeq  
\min_{\scs q_{XY} \in \mathcal{G}_2: 
     \atop{ \scs 
         {\rm E}_{q_X}[X^2]\leq \Gamma
     }
 }
\big\{ 
[R-I(q_X,q_{Y|X})]^+ + D(q_{Y|X}|| W | q_X)
\big\}, 
\end{align*}
where $[t]^+ = \max\{0,t\}$ and 
\begin{align*}
I(q_X, q_{Y|X}) 
& =
{\rm E}_q \left[
\log \frac{q_{Y|X}(Y|X)}{q_{Y}(Y)}
\right], \\
D(q_{Y|X} || W | q_X)
& = 
\mathrm{E}_q \left[\log \frac{q_{Y|X}(Y|X)}{W(Y|X)}
\right].
\end{align*}
Then we have the following theorem. 
\begin{Th} \label{Th:GauDK}
For any $R>0$, 
\begin{align*}
      & G^*(R,\Gamma |{\NoizeVar}) \leq  
\overline{G}_{\rm DK}(R,\Gamma|{\NoizeVar}).
\notag
\end{align*}
\end{Th}

Proof of this theorem is given in Appendix \ref{sub:ApdaDirectDK}.
\newcommand{\ApdaDirectDK}{
\subsection{Proof of Theorem \ref{Th:GauDK}} 
\label{sub:ApdaDirectDK}

We fix $\delta \in [0,1/2)$. We consider a Gaussian channel 
with input $X$ and output $Y$, having the form 
$$
Y=\alpha X+ S, X\perp S, 
S\sim {\cal N}(0,\xi).
$$
We consider a Gaussian random pair  $(X,Y)$ obtained by letting $X$ be 
Gaussian random variable with $X\sim {\cal N}(0,\theta)$. We assume that 
$\theta \leq \Gamma$. Let $q_{XY}$ be a probability density function of 
$(X,Y)$. For the Gaussian channel specified by $q_{Y|X}$, we can 
construct an $n$-length block code $(\phi^{(n)}, \psi^{(n)})$ with 
message set ${\cal K}_n$ satisfying:
\begin{itemize}
\item[a)] ${\rm P}_{\rm c}^{(n)}(\phi^{(n)}, \psi^{(n)}|
q_{Y|X})\geq 1-\delta$.
\item[b)] all codewords $\phi^{(n)}(k), k\in {\cal K}_n$ satisfy 
$||\phi^{(n)}(k)||^2 \leq n \theta$. 
\item[c)] $\frac1n \log |\mathcal{ K }_n| 
\geq \min \{ R, I(q_X,q_{Y|X}) -\delta \}$.
\end{itemize}
By the condition b), we can obtain the following result. 
\begin{lm}\label{lm:Aszz}
For every $k\in {\cal K}_n$, we have 
\begin{align}
& \underbrace{\int \cdots \int}_{n}{\rm d}y^n
q_{Y|X}^n(y^n|\phi^{(n)}(k)) \log 
\frac{q_{Y|X}^n(y^n|\phi^{(n)}(k))}{W^n(y^n|\phi^{(n)}(k))}
\notag\\
&\leq n D(q_{Y|X}||W|q_X) 
\label{eq.zzz6}.
\end{align}
\end{lm}

{\it Proof:} By a direct computation we have 
\begin{align}
& \underbrace{\int \cdots \int}_{n} {\rm d}y^n
q_{Y|X}^n(y^n|\phi^{(n)}(k)) \log 
\frac{q_{Y|X}^n(y^n|\phi^{(n)}(k))}{W^n(y^n|\phi^{(n)}(k))}
\notag\\
&=
\frac{1}{2}(1-\alpha)^2
\frac{||\phi^{(n)}(k))||^2}{{\NoizeVar}}
+\frac{n}{2}\left[\frac{\xi}{{\NoizeVar}}-1+\log \frac{{\NoizeVar}}{\xi}\right]
\notag\\
&\MLeq{a}
\frac{n}{2}(1-\alpha)^2 \frac{\theta}{{\NoizeVar}}
+\frac{n}{2}\left[\frac{\xi}{{\NoizeVar}}-1+\log \frac{{\NoizeVar}}{\xi}\right]
\notag\\
&=nD(q_{Y|X} || W | q_X). 
\end{align}
Step (a) follows from the condition b).
\hfill\IEEEQED
%

For $k\in {\cal K}_n$, we set
\beqno
\alpha_n(k)&\defeq&W^n( {\mathcal{D}(k)}|\phi^{(n)}(k))=
\sum_{y^n \in \mathcal{D}(k)} W^n(y^n|\phi^{(n)}(k)),
\\
\beta_n(k)&\defeq&q_{Y|X}^n( {\mathcal{D}(k)}|\phi^{(n)}(k))=
\sum_{y^n \in \mathcal{D}(k)} q_{Y|X}^n(y^n|\phi^{(n)}(k)),
\\
\overline{\alpha_n(k)}&\defeq&
1-{\alpha_n(k)}=q_{Y|X}^n(\overline{\mathcal{D}(k)}|\phi^{(n)}(k)),
\\
\overline{\beta_n(k)}&\defeq& 
1-{\beta_n(k)}=q_{Y|X}^n(\overline{\mathcal{D}(k)}|\phi^{(n)}(k)).
\eeqno
Furthermore, set 
\beqno
&&{\alpha_n}\defeq\sum_{k\in {\cal K}_n}
 \frac{1}{|{\cal K}_n|} \alpha_n(k)
={\rm P}_{\rm c}^{(n)}(\phi^{(n)}, \psi^{(n)}|{W}),
\\
&&{\beta_n}\defeq\sum_{k\in {\cal K}_n}
 \frac{1}{|{\cal K}_n|} \beta_n(k)
={\rm P}_{\rm c}^{(n)}(\phi^{(n)}, \psi^{(n)}|q_{Y|X}).
\eeqno
The quantity 
${\rm P}_{\rm c}^{(n)}(\phi^{(n)}, \psi^{(n)}|{W})$
has a lower bound given by the following Lemma. 
\begin{lm}\label{lm:DirectLm} For any $\delta\in [0,1/2)$, we have
\begin{align}
& {\rm P}_{\rm c} ^{(n)}( \phi^{(n)}, \psi^{(n)}|{W})
=\frac1{|{\mathcal{K}}_n|}
\sum_{k \in \mathcal{K}_n} W^n(\mathcal{D}(k)|\phi^{(n)}(k)) 
\notag\\
&\geq \exp \{-n [(1-\delta)^{-1}D(q_{Y|X}||W |q_X) + \eta_n(\delta)]\}.
\label{eqn:Saab} 
\end{align}
Here we set
$
\eta_n(\delta) \defeq \frac 1 n (1-\delta)^{-1}h(1 -\delta) 
$
and $h(\cdot)$ stands for a binary entropy function.
\end{lm}

{\it Proof:} 
We have the following chain of inequalities: 
\beqa
& &nD(q_{Y|X} || W | q_X) 
\nonumber\\
&\MEq{a}&\frac{1} {|\mathcal{K}_n|} \sum_{k\in \mathcal{K}_n} 
\sum_{y^n \in \mathcal{Y}^n} 
q_{Y|X}^n(y^n|\phi^{(n)}(k)) \log 
\frac{q_{Y|X}^n(y^n|\phi^{(n)}(k))}{W^n(y^n|\phi^{(n)}(k))}
\nonumber\\
&\MGeq{b} &
\frac{1} {|\mathcal{K}_n|} \sum_{k\in \mathcal{K}_n} \left[
\beta_n (k) \log \frac{\beta_n (k)}{\alpha_n(k)}
+\overline{\beta_n (k)} \log \frac{\overline{\beta_n (k)}}
{\overline{\alpha_n(k)}}\right]
\nonumber\\
&=&  \sum_{k\in \mathcal{K}_n} 
\left[\frac{\beta_n (k)}{|\mathcal{K}_n|}
\log \frac
{\frac{\beta_n (k)}{|\mathcal{K}_n|}}
{\frac{\alpha_n(k)}{|\mathcal{K}_n|}}
+\frac{\overline{\beta_n (k)}}{|\mathcal{K}_n|} 
\log \frac
{\frac{\overline{\beta_n (k)}}{|\mathcal{K}_n|}}
{\frac{\overline{\alpha_n(k)}}{|\mathcal{K}_n|}}
\right]
\nonumber\\
&\MGeq{c}&
\beta_n \log \frac{\beta_n}{\alpha_n}
+\overline{\beta_n} 
\log \frac{\overline{\beta_n}}{\overline{\alpha_n}}
\geq -h(\beta_n) -\beta_n\log \alpha_n
\notag \\
&\MGeq{d}& -h(1-\delta)-(1-\delta)\log \alpha_n. 
\label{eqn:Zsddv}
\eeqa
Step (a) follows from Lemma \ref{lm:Aszz}. Steps (b) and (c) 
follow from the log-sum inequality. Step (d) follows from that 
$$
\beta_n={\rm P}_{\rm c} ^{(n)}(\phi^{(n)}, \psi^{(n)}|q_{Y|X})\geq 1-\delta
$$
and $\delta \in  (0,1/2].$  From (\ref{eqn:Zsddv}), we obtain   
\begin{align}
& \alpha_n= {\rm P}_{\rm c} ^{(n)}( \phi^{(n)}, \psi^{(n)}|{W})
\notag\\
&\geq \exp\left(
- \frac{nD(q_{Y|X}|| W| q_X) + h(1-\delta)}
{1-\delta} \right)
\notag\\
&=\exp \{-n [(1-\delta)^{-1}D(q_{Y|X}||W|q_X) + \eta_n(\delta)]\},
\notag
\end{align}
completing the proof. \hfill\IEEEQED 

{\it Proof of Theorem \ref{Th:GauDK}:} 
We first consider the case where 
$R \leq I(q_X, q_{Y|X}) - \delta$. 
In this case we choose $\varphi^{(n)}=\phi^{(n)}$. Then we have 
\begin{align}
& {\rm P}_{\rm c} ^{(n)}( \varphi^{(n)}, \psi^{(n)}|W)
  ={\rm P}_{\rm c} ^{(n)}( \phi^{(n)}, \psi^{(n)}|W)
\notag\\
&\MEq{a}\exp \{ -n[R+\delta-I(q_X,q_{Y|X})]^+
\notag\\
&\quad -n [(1-\delta)^{-1}D(q_{Y|X} || W | q_X) + \eta_n(\delta)]\}
\notag\\
&\MGeq{b}\exp \{-n[R-I(q_X,q_{Y|X})]^+
\notag\\
&\quad -n [(1-\delta)^{-1}D(q_{Y|X} || W | q_X) + \delta+ \eta_n(\delta)]\}.
\label{eqn:SaabZ}
\end{align}
Step (a) follows from the condition 
$R+\delta-I(q_X,q_{Y|X})\leq 0$. 
Step (b) follows from that
$$
[R+\delta-I(q_X,q_{Y|X})]^+ \leq [R-I(q_X,q_{Y|X})]^+ +\delta.
$$
We next consider the case where $R>I(q_X,q_{Y|X})-\delta$. 
Consider the new message set $\widehat{ \mathcal{K}}_n$ satisfying 
$|\widehat{\mathcal{K}}_n |={\rm e}^{\lfloor nR \rfloor}$. 
For new message set $\widehat{\cal K}_n$, we define 
$\varphi^{(n)}(k)$ such that 
$\varphi^{(n)}(k) = \phi^{(n)}(k)$ if $k\in \mathcal{K}_n$.
For $k \in \widehat{\mathcal{K}}_n-\mathcal{K}_n$, 
we define $\varphi^{(n)}(k)$
arbitrary sequence of ${\cal X}^n$ having the type $q_X$. 
We use the same decoder $\psi^{(n)}$ 
as that of the message set $\mathcal{K}_n$.
Then we have the following: 
\begin{align}
&{\rm P}_{\rm c} ^{(n)}(\varphi^{(n)}, \psi^{(n)}|\NoizeVar)
\notag\\
&=
\frac{1}{ | \widehat{\mathcal{K} }_n |}
\left[\sum_{k\in \mathcal{K}_n } W^n(\mathcal{D}(k)|\varphi^{(n)}(k)) 
\right. \notag\\
&\qquad \left. + 
\sum_{k\in \widehat{\mathcal{K}}_n - \mathcal{K}_n} 
W^n(\mathcal{D}(k)|\varphi^{(n)}(k))\right] 
\notag\\
& \geq 
\frac 1 {|\widehat{\mathcal{K}}_n|}
\sum_{k\in \mathcal{K}_n} W^n(\mathcal{D}(k)|\varphi^{(n)}(k)) 
\notag\\
& \MGeq{a}
\frac {|\mathcal{K}_n|}{{\rm e}^{nR}} 
\exp\{ -n [(1-\delta)^{-1}D(q_{Y|X}||W|q_{X}) +\eta_n({\delta})]\}
\notag\\
&\MGeq{b} \exp \left[
-n \left\{R-(I(q_X, q_{Y|X})- \delta)
\right.\right. 
\notag\\
&\left.\left.\quad\qquad+ (1-\delta)^{-1}D(q_{Y|X}||W |q_X) 
+ \eta_n(\delta)\right\} \right]
\notag\\
&\MGeq{c} \exp \left[-n \left\{
[R-I(q_X, q_{Y|X})]^+ \right.\right.
\notag\\
&\left.\left.\quad \qquad +(1-\delta)^{-1}D(q_{Y|X}||W|q_X)
+\delta+ \eta_n(\delta)\right\}\right].
\label{eqn:eq29}
\end{align}
Step (a) follows from (\ref{eqn:Saab}) in Lemma \ref{lm:DirectLm}. 
Step (b) follows from 
$|\mathcal{K}_n| \geq {\rm e}^{n [(I(q_X, q_{Y|X})-\delta]}.$
Step (c) follows from $[a]\leq [a]^+$. 
Combining (\ref{eqn:SaabZ}) and (\ref{eqn:eq29}), we have
\beqa
& &{\rm P}_{\rm c} ^{(n)}( \varphi^{(n)}, \psi^{(n)}|W)
\nonumber\\
&\geq& \exp \left[-n \left\{
[R-I(q_X, q_{Y|X})]^+ \right.\right.
\nonumber\\
& &\left.\left.\quad +(1-\delta)^{-1}D(q_{Y|X}||W|q_X)
+\delta+ \eta_n(\delta)\right\}\right]
\label{eq.30}
\eeqa
for any $q_X \in {\cal G}_1$ with 
${\rm E}_{q_X}[X^2] \leq \Gamma$ and 
$q_{Y|X} \in {\cal G}_1($$q_X)$.
Hence from (\ref{eq.30}), we have
\begin{align}
&- \frac1 n \log {\rm P}_{\rm c} ^{(n)}( \varphi^{(n)}, \psi^{(n)}|\NoizeVar)
\notag\\
& \leq 
\min_{\scs (q_X,q_{Y|X}) \in \mathcal{G}_2, 
  \atop{\scs 
        {\rm E}_{q_X}[X^2] \leq \Gamma
  }
}
\{ [R -I(q_X, q_{Y|X}) ]^+ 
\notag\\
&\quad + (1-\delta)^{-1}D(q_{Y|X} || W | q_X) + \delta + \eta_n(\delta) \} 
\notag\\
& \leq  
(1-\delta)^{-1}\min_{\scs q_{XY} \in \mathcal{G}_2, 
  \atop{\scs 
        {\rm E}_{q_X}[X^2] \leq \Gamma
  }
}
\{[R -I(q_X, q_{Y|X}) ]^+ 
\notag\\
&\quad + D(q_{Y|X}||W|q_X)\}+\delta +\eta_n(\delta)
\notag\\
& \leq (1-\delta)^{-1}G_{\rm DK}(R,\Gamma|\NoizeVar)+\delta 
 + \eta_n(\delta).
\label{eqn:Bound}
\end{align}
We note that
$\eta_n(\delta)\to 0$ as $n\to\infty$. Hence by letting 
$n\to \infty$ in (\ref{eqn:Bound}), we obtain 
$$
G^*(R,\Gamma|\NoizeVar) \leq (1-\delta)^{-1}G_{\rm DK}(R,\Gamma|\NoizeVar)+\delta.
$$
Since $\delta$ can be made arbitrary small, 
we conclude that
$
G^*(R$ $,\Gamma|\NoizeVar) \leq G_{\rm DK}(R,\Gamma|\NoizeVar).
$
\hfill \IEEEQED
}
We next derive a lower bound of $G^*(R,\Gamma |{\NoizeVar})$. To this end 
we define several quantities. Define
\beqno
& &\Omega^{(\mu,\lambda)}(q_X,Q|{\NoizeVar})
\\
& & \defeq \log\left[\int \int {\rm d}x  {\rm d}y
q_{X}(x)
\frac{ {\rm e}^{-\frac{(1+\lambda)(y-x)^2}{2{\NoizeVar}}}
{\rm e}^{-\mu\lambda x^2}}{
({\sqrt{2\pi {\NoizeVar}}})^{1+\lambda}
Q^\lambda(y)}
\right],
\\
& &\Omega^{(\mu,\lambda)}({\NoizeVar})
\defeq 
\max_{q_X}
\min_{Q}
\Omega^{(\mu,\lambda)}(q_X,Q|{\NoizeVar}),
\\
& &
G_{\rmOH}^{(\mu,\lambda)}(R,\Gamma|{\NoizeVar})
\defeq
\frac{\lambda (R-\mu\Gamma) -\Omega^{(\mu, \lambda)}({\NoizeVar})}
{1+\lambda},
\\
& &G_{\rmOH}(R,\Gamma|{\NoizeVar})
\defeq \sup_{\mu,\lambda \geq 0} 
G_{\rmOH}^{(\mu,\lambda)}(R,\Gamma |{\NoizeVar}).
\eeqno

According to Oohama \cite{SingleStConvArXiv17}, 
we have the following theorem.
\begin{Th}[Oohama \cite{SingleStConvArXiv17}]
\label{Th:GauMain}  
For any AWGN $W$, we have 
\beqa
G^*(R,\Gamma|{\NoizeVar}) &\geq& G_{\rmOH}(R,\Gamma|{\NoizeVar}). 
\label{eqn:GaumainIeq}
\eeqa
\end{Th}
\newcommand{\prmt}{\lambda}

To find an explicit formula of $G_{\rmOH}(R,\Gamma|{\NoizeVar})$, set    
\beqno
& &\underline{\Omega}^{(\mu,\lambda)}({\NoizeVar})
\defeq 
\max_{q_X\in {\cal G}_1}
\min_{Q}
\Omega^{(\mu,\lambda)}(q_X,Q|{\NoizeVar}),
\\
& &\overline{\Omega}^{(\mu,\lambda)}({\NoizeVar})
\defeq 
\max_{q_X}
\min_{Q \in {\cal G}_1}
\Omega^{(\mu,\lambda)}(q_X,Q|{\NoizeVar}),
\\
& &
\overline{G_{\rmOH}}^{(\mu,\lambda)}(R,\Gamma|{\NoizeVar})
\defeq
\frac{\lambda (R-\mu\Gamma)-\underline{\Omega}^{(\mu, \lambda)}({\NoizeVar})}
{1+\lambda},
\\
& &
\underline{G_{\rmOH}}^{(\mu,\lambda)}(R,\Gamma|{\NoizeVar})
\defeq
\frac{\lambda (R-\mu\Gamma)-\overline{\Omega}^{(\mu, \lambda)}({\NoizeVar})}
{1+\lambda},
\\
& &\overline{G}_{\rmOH}(R,\Gamma|{\NoizeVar})
\defeq \sup_{\mu,\lambda \geq 0} 
\overline{G}_{\rmOH}^{(\mu,\lambda)}(R,\Gamma |{\NoizeVar}),
\\
& &\underline{G}_{\rmOH}(R,\Gamma|{\NoizeVar})
\defeq \sup_{\mu,\lambda \geq 0} 
\underline{G}_{\rmOH}^{(\mu,\lambda)}(R,\Gamma |{\NoizeVar}).
\eeqno
It is obvious that 
$$
\underline{G}_{\rmOH}(R,\Gamma|{\NoizeVar})
\leq {G}_{\rmOH}(R,\Gamma|{\NoizeVar})\leq 
\overline{G}_{\rmOH}(R,\Gamma|{\NoizeVar}).
$$
For $\lambda \in [0,1)$, define   
\beqno
& &{J}^{(\mu, \lambda)}(q_X|{\NoizeVar})
\\
& & \defeq \log\left\{
\int \!\!{\rm d}y \!
\left[
\int \!\!{\rm d}x 
q_X(x)\! \left\{
\GN{\rm e}^{-\mu \lambda x^2}\right\}^{\frac{1}{1-\lambda}}
\right]^{1-\lambda}\! \right\},
\\
& &{\empty}{G}_{\rm AR}^{(\mu,\lambda)}(R,\Gamma,q_X |{\NoizeVar}) 
\defeq \lambda (R-\mu \Gamma)-{J}^{(\mu, \lambda)}
(q_X|{\NoizeVar}),
\\
& &G_{\rm AR}^{(\mu,\prmt)}(R,\Gamma \Vl {\NoizeVar}) 
\defeq \min_{q_X}
{G}_{\rm AR}^{(\mu,\lambda)}(R,\Gamma, q_X \Vl {\NoizeVar}).
\eeqno
Furthermore, set
\beqno
& &{G}_{\rm AR}(R, \Gamma \Vl {\NoizeVar})
\defeq 
\sup_{\scs \mu \geq 0, \atop{\scs \lambda \in [0,1)}}
{G}_{\rm AR}^{(\mu, \lambda)}(R,\Gamma \Vl {\NoizeVar})
\\
&=& 
\sup_{
\scs \mu\geq 0, \atop{\scs \lambda \in [0,1)}}
\min_{q_X }
{G}_{\rm AR}^{(\lambda)}(R, \Gamma, q_X \Vl {\NoizeVar})
\\
&=&
\sup_{
\scs \mu\geq 0, \atop{\scs \lambda \in [0,1)}}
\left[
{\lambda (R-\mu\Gamma)}-\max_{q_X}
{J}^{(\mu,\lambda)}(q_X \Vl {\NoizeVar})
\right],
\\
& &\overline{G}_{\rm AR}(R, \Gamma \Vl {\NoizeVar})
\defeq 
\sup_{\scs \mu \geq 0, \atop{\scs \lambda \in [0,1)}}
\overline{G}_{\rm AR}^{(\mu, \lambda)}(R,\Gamma \Vl {\NoizeVar})
\\
&=& 
\sup_{
\scs \mu\geq 0, \atop{\scs \lambda \in [0,1)}}
\min_{q_X \in {\cal G}_1}
{\empty}{G}_{\rm AR}^{(\lambda)}(R, \Gamma, q_X \Vl {\NoizeVar})
\\
&=&
\sup_{
\scs \mu\geq 0, \atop{\scs \lambda \in [0,1)}}
\left[
{\lambda (R-\mu\Gamma)}-\max_{q_X \in {\cal G}_1}
{J}^{(\mu,\lambda)}(q_X \Vl {\NoizeVar})
\right].
\eeqno
According to Oohama \cite{SingleStConvArXiv17}, we have the following 
lemma.
\begin{lm}[Oohama \cite{SingleStConvArXiv17}]For any $q_X$, 
\label{lm:AssdZ} 
$$
\min_{Q}\Omega^{(\mu,\lambda)}(q_X,Q|{\NoizeVar})
=(1+\lambda)J^{(\mu,\frac{\lambda}{1+\lambda})}(q_X|{\NoizeVar}).
$$
The above minimization is attained by the probability density function $Q$ 
having the form 
\beq
Q(y)=\kappa \left[ 
\int q_X(x) \left\{
\frac{{\rm e}^{-\frac{(y-x)^2}{2{\NoizeVar}}}}
{\sqrt{2\pi {\NoizeVar}}}
\right\}^{1+\lambda}
{\rm e}^{-\mu \lambda x^2}
{\rm d}x
\right]^{\frac{1}{1+\lambda}},
\label{eqn:Assjj}
\eeq
where 
\beqa
\kappa^{-1}
&=& \int {\rm d}y
\left[
\int {\rm d}x 
q_X(x)\left\{
\GN{\rm e}^{-\mu \lambda x^2}\right\}^{\frac{1}{1-\lambda}}
\right]^{1-\lambda}
\nonumber\\
&=&\exp\left\{{J}^{(\mu, \frac{\lambda}{1+\lambda})}(q_X|{\NoizeVar})\right\}
\label{eqn:Azii}
\eeqa
is a constant for normalization. 
\end{lm}

From Lemma \ref{lm:AssdZ}, we have the following proposition.
\begin{pro}
\label{pro:proZ}
For any $\mu, \lambda \geq 0$, we have 
\beqno
& &{G}^{(\mu, \lambda)}(R,\Gamma \Vl {\NoizeVar})
   ={G}_{\rm AR}^{(\mu, \frac{\lambda}{1+\lambda})}(R,\Gamma \Vl {\NoizeVar}),
\\
& &\overline{G}^{(\mu, \lambda)}(R,\Gamma \Vl {\NoizeVar})
   =\overline{G}_{\rm AR}^{(\mu, \frac{\lambda}{1+\lambda})}
(R,\Gamma \Vl {\NoizeVar}).
\eeqno
In particular, we have 
\beqno
& &{G}(R,\Gamma \Vl {\NoizeVar})={G}_{\rm AR}(R,\Gamma \Vl {\NoizeVar}),
\\
& &\overline{G}(R,\Gamma \Vl {\NoizeVar})
=\overline{G}_{\rm AR}(R,\Gamma \Vl {\NoizeVar}).
\eeqno
\end{pro}

Let $q_{X,\theta}\in {\cal G}_1$ be the p.d.f. of the Gaussian 
distribution with mean 0 and variance $\theta$. We set 
\beqa
& &\xi=\xi(\mu,\lambda,\theta)\defeq \frac{(1+\lambda)\theta}{1+2\mu\lambda\theta}
=\frac{ 1+\lambda}{\ds \theta^{-1}+ 2\mu\lambda},
\label{eqn:DefXi}\\
& &\zeta^{(\mu,\lambda)}(\eta |{\NoizeVar})
\nonumber\\
& &\defeq
\frac{\lambda}{2}\log\left(1+\frac{ \eta }{{\NoizeVar}}\right)
+\frac{1}{2}\log\left( 
1-\frac{\lambda}{1+\lambda}\cdot 2\mu \eta \right),
\nonumber\\
& &L^{(\mu,\lambda)}(R,\Gamma|{\NoizeVar})
\nonumber\\
&&\defeq 
\frac{\lambda(R-\mu \Gamma)-\zeta^{(\mu,\lambda)}
\left(\frac{1}{2\mu}-\frac{{\NoizeVar}}{1+\lambda} 
\Bigl|{\NoizeVar}\right)}{1+\lambda} 
\nonumber\\
&&=\frac{\lambda}{1+\lambda}(R-\mu \Gamma)
-\frac{\lambda}{1+\lambda}
\frac{1}{2}\log
\left(\frac{\lambda}{1+\lambda}+\frac{1}{2\NoizeVar\mu}\right)
\nonumber\\
& &\quad -\frac{1}{1+\lambda}
\frac{1}{2}\log\left(\frac{1}{1+\lambda}
\left[1+\frac{2\mu\lambda {\NoizeVar}}{1+\lambda}\right]\right).
\eeqa
The following proposition is a key result of this paper. 
\begin{pro}\label{pro:proDss} 
For every $\lambda \geq 0$ and for every $ \mu 
\in [0, \frac{1+\lambda}{2{\NoizeVar}}]$, we have 
\beqa
& &\underline{G}^{(\mu, \lambda)}_{\rmOH}(R,\Gamma|{\NoizeVar})
={G}^{(\mu, \lambda)}_{\rmOH}(R,\Gamma|{\NoizeVar})
=\overline{G}^{(\mu, \lambda)}_{\rmOH}(R,\Gamma|{\NoizeVar})
\nonumber\\
&=&\overline{G}^{(\mu, \frac{\lambda}{1+\lambda})}_{\rm AR}
(R,\Gamma|{\NoizeVar})
=L^{(\mu,\lambda)}(R,\Gamma|{\NoizeVar}).
\label{eqn:zSSS}
\eeqa
The input Gaussian distribution $q_{X,\theta}\in {\cal G}_1$
attaining 
$\overline{G}_{\rm AR}^{(\mu, \frac{\lambda}{1+\lambda})}($ $R,\Gamma|\NoizeVar)$
satisfies the following:
$$
\xi=\xi(\theta)= \frac{(1+\lambda)\theta}{1+2\mu\lambda\theta}
=\frac{1}{2\mu}-\frac{{\NoizeVar}}{1+\lambda}.
$$
Furthermore, we have 
\beqa
& &{G}_{\rmOH}(R,\Gamma|{\NoizeVar})
=\max_{\mu,\lambda\geq 0}
{G}^{(\mu, \lambda)}_{\rmOH}(R,\Gamma|{\NoizeVar})
\nonumber\\
& &=\max_{\scs \lambda \geq 0,
\atop{\scs
\mu \in [0, \frac{1+\lambda}{2{\NoizeVar}}]}}
L^{(\mu,\lambda)}(R,\Gamma|{\NoizeVar}).
\label{eqn:Asxx}
\eeqa
\end{pro}

Proof of this proposition is given in Section \ref{sec:ThreeExp}. 
\newcommand{\LemmaForProposition}{

For $\theta \geq 0$, we let $\xi=\xi(\mu,\lambda,\theta)$ 
be the same quantity as that defined by (\ref{eqn:DefXi}).
Let $Q_{\xi+{\NoizeVar}} \in {\cal G}_1$ be a Gaussian distribution 
with mean 0 and variance $\xi+{\NoizeVar}$. Applying Lemma \ref{lm:AssdZ} to 
$q_{X,\theta} \in {\cal G}_1$, we have the following. 
\begin{lm}\label{lm:lmSdd} 
The function $\Omega^{(\mu,\lambda)}(q_{X,\theta},Q|{\NoizeVar}) $
takes the minimum value at $Q=Q_{\xi+{\NoizeVar}}$. The value is   
\begin{align}
&\min_{Q} \Omega^{(\mu,\lambda)}(q_{X,\theta},Q|{\NoizeVar})
=\Omega^{(\mu,\lambda)}(q_{X,\theta},Q_{\xi+{\NoizeVar}}|{\NoizeVar})
\notag\\
&=(1+\lambda)J^{(\mu,\frac{\lambda}{1+\lambda})}(q_{X,\theta}|{\NoizeVar})
=\zeta^{(\mu,\lambda)}(\xi(\mu,\lambda,\theta)|{\NoizeVar}).
\notag
\end{align}
Furthermore, we have
\begin{align}
& \underline{\Omega}^{(\mu,\lambda)}_{\rmOH}({\NoizeVar})
= \max_{0\leq \xi<\frac{1+\lambda}{2\mu\lambda}}
\zeta^{(\mu,\lambda)}(\xi|{\NoizeVar}),
\notag\\
& \overline{G}_{\rmOH}^{(\mu,\lambda)}(R,\Gamma|{\NoizeVar})
=  
\frac{\ds \lambda(R-\mu \Gamma)-
\max_{0\leq \xi<\frac{1+\lambda}{2\mu\lambda}}
\zeta^{(\mu,\lambda)}(\xi|{\NoizeVar})}{1+\lambda}.
\notag
\end{align}
\end{lm} 

{\it Proof: } By an elementary computation we have 
\begin{align}
&q_{X,\theta}(x)
\left\{
\frac{{\rm e}^{-\frac{(y-x)^2}{2{\NoizeVar}}}}
{{\sqrt{2\pi {\NoizeVar}}}}\right\}^{1+\lambda}
{\rm e}^{-\mu \lambda x^2}
\notag\\
&=
\frac{
{\rm e}^{-\frac{x^2}{2\theta}}}
{{\sqrt{2\pi \theta }}}
\frac
{ {\rm e}^{-\frac{(1+\lambda)(y-x)^2}{2{\NoizeVar}}}}
{({\sqrt{2\pi {\NoizeVar}}})^{1+\lambda}}
{\rm e}^{-\mu \lambda x^2}
\notag\\
&=\frac{{\rm e}^{-\frac{(1+\lambda)(\xi+{\NoizeVar})}{2\xi {\NoizeVar}}
\left(x-\frac{\xi}{\xi+{\NoizeVar}}y\right)^2}}
  {\sqrt{2\pi \theta }(\sqrt{2\pi {\NoizeVar}})^{1+\lambda} }
{\rm e}^{-\frac{(1+\lambda)y^2}{2(\xi+{\NoizeVar})}},
\label{eqn:Asssjj}
\end{align}
which together with Lemma \ref{lm:AssdZ} 
and (\ref{eqn:Assjj}) in this lemma yields that
$
\Omega^{(\mu,\lambda)}(q_{X,\theta},Q|{\NoizeVar})  
$
takes the minimum value at the p.d.f. $Q=Q(y)$ given by
\begin{align}
&Q(y)
=\kappa \left[ 
\int {\rm d}x q_{X,\theta}(x) \left\{
\frac{{\rm e}^{-\frac{(y-x)^2}{2{\NoizeVar}}}}
{\sqrt{2\pi {\NoizeVar}}}
\right\}^{1+\lambda}
{\rm e}^{-\mu \lambda x^2}
\right]^{\frac{1}{1+\lambda}}
\notag\\
&=\kappa \left[ 
\int {\rm d}x
\frac{{\rm e}^{-\frac{(1+\lambda)(\xi+{\NoizeVar})}{2\xi {\NoizeVar}}
\left(x-\frac{\xi}{\xi+{\NoizeVar}}y\right)^2}}
  {\sqrt{2\pi \theta } }
\right]^{\frac{1}{1+\lambda}} 
\frac{{\rm e}^{-\frac{y^2}{2(\xi+{\NoizeVar})}}}{\sqrt{2\pi {\NoizeVar}}}
\notag\\
&=\kappa \left[\frac{\xi {\NoizeVar}}{(1+\lambda)\theta(\xi+{\NoizeVar})}
          \right]^{\frac{1}{2(1+\lambda)}}
   \left[1+\frac{\xi}{{\NoizeVar}}\right]^{\frac{1}{2}}
\notag\\
& \qquad \times \frac{{\rm e}^{-\frac{(1+\lambda)y^2}{2(\xi+{\NoizeVar})}}}
          {\sqrt{2\pi(\xi+{\NoizeVar})}}=Q_{\xi+{\NoizeVar}}(y).
\label{eqn:asQQcc}
\end{align}
From (\ref{eqn:asQQcc}), we can see that $\kappa$ must satisfy
\beq
\kappa^{-1}=
\left[\frac{\xi {\NoizeVar}}{(1+\lambda)\theta(\xi+{\NoizeVar})}
\right]^{\frac{1}{2(1+\lambda)}}
\times \left[1+\frac{\xi}{{\NoizeVar}}\right]^{\frac 1 2}.
\label{eqn:asQQcc2}
\eeq
The minimum value is 
\beqa
& &(1+\lambda){J}^{(\mu, \frac{\lambda}{1+\lambda})}(q_X|{\NoizeVar})
 \MEq{a}(1+\lambda)\log\left(\kappa^{-1}\right)
\nonumber\\
&\MEq{b}&\frac{1}{2}\log 
\left[\frac{\xi {\NoizeVar}}{(1+\lambda)\theta(\xi+{\NoizeVar})}\right]
+\frac{1+\lambda}{2}\log \left(1+\frac{\xi}{{\NoizeVar}}\right)
\nonumber\\
&\MEq{c}&\frac{1}{2}\log
\left [\frac{\xi {\NoizeVar}}{\xi+{\NoizeVar}}
\left(\frac{1}{\xi}-\frac{2\lambda\mu}{1+\lambda} \right)\right]
+\frac{1+\lambda}{2}\log \left(1+\frac{\xi}{{\NoizeVar}}\right)
\nonumber\\
&=&
\frac{\lambda}{2}\log\left(1+\frac{\xi}{{\NoizeVar}}\right)
+\frac{1}{2} \log
\left(1-\frac{\lambda}{1+\lambda}\cdot 2\mu\xi \right).
\nonumber
\eeqa
Step (a) follows from (\ref{eqn:Azii}) in Lemma \ref{lm:AssdZ}.
Step (b) follows from (\ref{eqn:asQQcc2}).
Step (c) follows from that by the definition of $\xi$, we have 
$$
\frac{1}{(1+\lambda)\theta}=\frac{1}{\xi}-\frac{2\lambda\mu}{1+\lambda}.
$$
Hence we have 
\begin{align}
& \underline{\Omega}^{(\mu,\lambda)}_{\rmOH}({\NoizeVar})
=\max_{\theta \geq 0 }\zeta^{(\mu,\lambda)}(\xi(\mu,\lambda, \theta)|{\NoizeVar})
\notag\\
&
=\max_{\scs \theta \geq 0, 
\atop{\xi=\frac{(1+\lambda)\theta}{1+2\mu\lambda\theta}
      }
}\zeta^{(\mu,\lambda)}(\xi|{\NoizeVar})
=\max_{0\leq \xi<\frac{1+\lambda}{2\mu\lambda}}
\zeta^{(\mu,\lambda)}(\xi|{\NoizeVar}),
\notag\\
& \overline{G}_{\rmOH}^{(\mu,\lambda)}(R,\Gamma|{\NoizeVar})
=  
\frac{\ds \lambda(R-\mu \Gamma)-
\max_{0\leq \xi<\frac{1+\lambda}{2\mu\lambda}}
\zeta^{(\mu,\lambda)}(\xi|{\NoizeVar})}
{1+\lambda},
\notag
\end{align}
completing the proof. 
\hfill \IEEEQED

On a lower bound of 
$\underline{G}_{\rmOH}(R,\Gamma|{\NoizeVar})$, 
we have the following lemma. 
\begin{lm}\label{lm:Aggd}
Suppose that $\mu \in [0, \frac{1+\lambda}{2{\NoizeVar}}]$. 
We choose $\eta=\eta(\mu,\lambda)$ so that
\beq
\eta=\eta(\mu,\lambda)=\frac{1}{2\mu}-\frac{{\NoizeVar}}{1+\lambda}.
\label{eqn:Awwwz}
\eeq 
Then for any $q_X$, we have
\begin{align}
&\min_{Q\in {\cal G}_1} \Omega^{(\mu,\lambda)}(q_X,Q|{\NoizeVar})
\leq \Omega^{(\mu,\lambda)}(q_X,Q_{\eta+{\NoizeVar}}|{\NoizeVar})
\notag\\
&=\zeta^{(\mu,\lambda)}\left(\eta |{\NoizeVar}\right)
 =\zeta^{(\mu,\lambda)}
  \left(\frac{1}{2\mu}-\frac{{\NoizeVar}}{1+\lambda}\Big| {\NoizeVar}\right)
\notag\\
&=
\frac{1}{2}\log
\left(\frac{\lambda}{1+\lambda}+\frac{1}{2{\NoizeVar}\mu}\right)
\notag\\
&\quad +\frac{1}{2}\log\left(\frac{1}{1+\lambda}
\left[1+\frac{2\mu\lambda {\NoizeVar}}{1+\lambda}\right]\right).
\notag
\end{align}
This implies that
\begin{align}
& \overline{\Omega}^{(\mu,\lambda)}_{\rmOH}({\NoizeVar})
  \leq \max_{q_X}{\Omega}^{(\mu,\lambda)}(q_X,Q_{\eta+{\NoizeVar}}|{\NoizeVar})
\notag\\
& = \zeta^{(\mu,\lambda)}
  \left(\frac{1}{2\mu}-\frac{{\NoizeVar}}{1+\lambda}\Big| {\NoizeVar}\right),
\notag\\
& \underline{G}_{\rmOH}^{(\mu,\lambda)}(R,\Gamma|{\NoizeVar})
\geq L^{(\mu,\lambda)}(R,\Gamma|{\NoizeVar}).
\notag
\end{align}
\end{lm}

{\it Proof:} 
By an elementary computation we have 
\begin{align}
&
\frac{{\rm e}^{-\frac{(y-x)^2}{2{\NoizeVar}}}}
{\sqrt{2\pi {\NoizeVar}}}
\left\{
\frac{{\rm e}^{-\frac{(y-x)^2}{2{\NoizeVar}}}}
{{\sqrt{2\pi {\NoizeVar}}}Q_{\eta+{\NoizeVar}}(y)}
\right\}^{\lambda}
{\rm e}^{-\mu \lambda x^2}
\notag\\
&=\frac{
\left(1+\frac{\eta}{{\NoizeVar}} \right)^{\frac{\lambda}{2}}}
{\sqrt{2\pi {\NoizeVar}}}
{\rm e}^{
-\frac{(1+\lambda)}{2{\NoizeVar}}(y-x)^2
+\frac{y^2}{2(\eta+{\NoizeVar})}-\mu \lambda x^2
}
\notag\\
&=
\frac{
\left(1+\frac{\eta}{{\NoizeVar}} \right)^{\frac{\lambda}{2}}}
{\sqrt{2\pi {\NoizeVar}}}
{\rm e}^{
-\frac{1}{2}
\cdot \frac{(1+\lambda)\eta+{{\NoizeVar}}
}{{\NoizeVar}(\eta+{\NoizeVar})}
\left(y-\frac{\eta+{\NoizeVar}}{\eta+\frac{{\NoizeVar}}{1+\lambda}}x\right)^2
}
\notag\\
&\quad \times
{\rm e}^{
\lambda \left(\frac{1}{\eta+ \frac{{\NoizeVar}}{1+\lambda}}-2\mu\right)x^2
}
%
\notag\\
&\MEq{a}\frac{
\left(1+\frac{\eta}{{\NoizeVar}} \right)^{\frac{\lambda}{2}}}
{\sqrt{2\pi {\NoizeVar}}}
{\rm e}^{
-\frac{1}{2}
\cdot \frac{(1+\lambda)\eta+{{\NoizeVar}}}{{\NoizeVar}(\eta+{\NoizeVar})}
\left(y-\frac{\eta+{\NoizeVar}}{\eta+\frac{{\NoizeVar}}{1+\lambda}}x\right)^2
}.
\label{eqn:Asssvv}
\end{align}
Step (a) follows from (\ref{eqn:Awwwz}). Then we have 
the following chain of equalities:
\begin{align}
&\exp \left\{
\Omega^{(\mu,\lambda)}(q_{X},Q_{\eta+{\NoizeVar}}|{\NoizeVar}) 
  \right\}
\notag\\
&=\int \!\! \int \!{\rm d}x{\rm d}y
q_{X}(x)
\frac{{\rm e}^{-\frac{(y-x)^2}{2{\NoizeVar}}}}
{\sqrt{2\pi {\NoizeVar}}}
\left\{
\frac{{\rm e}^{-\frac{(y-x)^2}{2{\NoizeVar}}}}
{{\sqrt{2\pi {\NoizeVar}}}Q_{\eta+{\NoizeVar}}(y)}
\right\}^{\lambda}
{\rm e}^{-\mu \lambda x^2}
\notag\\
&\MEq{a}
\frac{\left(1+\frac{\eta}{{\NoizeVar}} \right)^{\frac{\lambda}{2}}}
{\sqrt{2\pi {\NoizeVar}}}
\int \! {\rm d}x
q_{X}(x)
\notag\\
&\qquad \times \int \! {\rm d}y 
{\rm e}^{
-\frac{1}{2}
\cdot \frac{(1+\lambda)\eta+{{\NoizeVar}}}{{\NoizeVar}(\eta+{\NoizeVar})}
\left(y-\frac{\eta+{\NoizeVar}}{\eta+\frac{{\NoizeVar}}{1+\lambda}}x\right)^2
}
\notag\\
&=\frac{\left(1+\frac{\eta}{{\NoizeVar}} \right)^{\frac{\lambda}{2}}}
{\sqrt{2\pi {\NoizeVar}}}
\int \! {\rm d}x
q_{X}(x)\cdot \sqrt{2\pi\cdot \frac{{\NoizeVar}(\eta+{\NoizeVar})}{(1+\lambda)\eta+{\NoizeVar}}}
\notag\\
&=\left(1+\frac{\eta}{{\NoizeVar}} \right)^{\frac{\lambda}{2}}
   \sqrt{\frac{\eta+{\NoizeVar}}{(1+\lambda)\eta+{\NoizeVar}}}.
\notag
\end{align}
Step (a) follows from (\ref{eqn:Asssvv}). Hence we have
\begin{align}
&\Omega^{(\mu,\lambda)}(q_{X},Q_{\eta+{\NoizeVar}}|{\NoizeVar}) 
\notag\\
&=\frac{\lambda}{2}\log\left(1+\frac{\eta}{{\NoizeVar}}\right)
+\frac{1}{2}\log\left(\frac{\eta+{\NoizeVar}}{(1+\lambda)\eta+{\NoizeVar}}
\right)
\notag\\
&=\frac{\lambda}{2}\log\left(1+\frac{\eta}{{\NoizeVar}}\right)
+\frac{1}{2}\log
\left(1-\frac{\lambda\eta}{(1+\lambda)\eta+{\NoizeVar}}\right)
\notag\\
&\MEq{a}\frac{\lambda}{2}\log\left(1+\frac{\eta}{{\NoizeVar}}\right)
+\frac{1}{2}\log\left( 
1-\frac{\lambda}{1+\lambda}\cdot 2\mu\eta \right)
\notag\\
&=\zeta^{(\mu,\lambda)}\left(\eta |{\NoizeVar}\right)
 \MEq{b}\zeta^{(\mu,\lambda)}
  \left(\frac{1}{2\mu}-\frac{{\NoizeVar}}{1+\lambda}\Big| {\NoizeVar}\right).
\notag
\end{align}
Step (a) follows from that by (\ref{eqn:Awwwz}), we have 
$$
\frac{1}{(1+\lambda)\eta+{\NoizeVar}}= \frac{\mu}{1+\lambda}.
$$
Step (b) follows from (\ref{eqn:Awwwz}).
\hfill\IEEEQED
\newcommand{\eWa}{
\begin{align}
&\frac{\lambda}{2}\log\left(1+\frac{\eta}{{\NoizeVar}}\right)
-\frac{1}{2}\log
\left(1-\frac{\lambda}{1+\lambda}\cdot 2\mu\eta \right),
\notag\\
&\max_{\scs \lambda\geq 0, \mu \in [0,\frac{1+\lambda}{2{\NoizeVar}}],
 \atop{ \scs 
 \eta=\frac{1}{2\mu}-\frac{{\NoizeVar}}{1+\lambda}
 }} 
\biggl\{ 
\frac{\lambda}{1+\lambda}(R-\mu \Gamma)-\frac{1}{1+\lambda}
\notag\\
&\times 
\max_{\eta\geq 0 }
\left[
\frac{\lambda}{2}\log\left(1+\frac{\eta}{{\NoizeVar}}\right)
-\frac{1}{2} \log
\left(1-\frac{\lambda}{1+\lambda}\cdot 2\mu\eta \right)
\right]
\biggr\}.
\end{align}
}
}
\newcommand{\ProofPropositionA}{

{\it Proof of Proposition \ref{pro:proDss}:} We assume that 
$\lambda \geq 0$ and $ \mu \in [0, \frac{1+\lambda}{2{\NoizeVar}}]$.
In this case by Lemma \ref{lm:lmSdd}, we have
\begin{align}
&\overline{G}_{\rmOH}^{(\mu,\lambda)}(R,\Gamma|{\NoizeVar})
\notag\\
&\leq 
\frac{\lambda(R-\mu \Gamma)-
\zeta^{(\mu,\lambda)}
\left(\frac{1}{2\mu}-\frac{{\NoizeVar}}{1+\lambda}\Big| {\NoizeVar}\right)
}{1+\lambda}
\notag\\
& = L^{(\mu,\lambda)}(R,\Gamma|{\NoizeVar}),
\label{eqn:Azzp0}
\end{align}
which together with Lemma \ref{lm:Aggd} yields 
the equality (\ref{eqn:zSSS}). We next prove (\ref{eqn:Asxx}).
We set
\beqno
A\defeq \max_{\scs \lambda \geq 0,
\atop{\scs
\mu \in [0, \frac{1+\lambda}{2{\NoizeVar}}]}}
G_{\rmOH}^{(\mu,\lambda)}(R,\Gamma|{\NoizeVar}),
\\
B\defeq
\max_{\scs \lambda \geq 0,
\atop{\scs
\mu \in [\frac{1+\lambda}{2{\NoizeVar}},\infty)}}
G_{\rmOH}^{(\mu,\lambda)}(R,\Gamma|{\NoizeVar}).
\eeqno
Then we have 
\begin{align}
&{G}_{\rmOH}(R,\Gamma|{\NoizeVar})
 =\max_{\lambda,\mu \geq 0}
G_{\rmOH}^{(\mu,\lambda)}(R,\Gamma|{\NoizeVar})
=\max\left\{A,B\right\}.
\end{align}
By the equality (\ref{eqn:zSSS}) already proved, we have
\beqa
A&=& \max_{\scs \lambda \geq 0,
\atop{\scs
\mu \in [0, \frac{1+\lambda}{2{\NoizeVar}}]}}
G_{\rmOH}^{(\mu,\lambda)}(R,\Gamma|{\NoizeVar})
\nonumber\\
&=&\max_{\scs \lambda \geq 0,
\atop{\scs
\mu \in [0, \frac{1+\lambda}{2{\NoizeVar}}]}}
L^{(\mu,\lambda)}(R,\Gamma|{\NoizeVar}).
\label{eqn:Asddvv}
\eeqa
Hence it suffices to prove $B\leq A$. 
When $\lambda\geq 0$ and $\mu \geq \frac{1+\lambda}{2{\NoizeVar}}$, 
we have the following chain of 
inequalities:
\begin{align}
& {G}_{\rmOH}^{(\mu,\lambda)}(R,\Gamma|{\NoizeVar})
\leq \overline{G}_{\rmOH}^{(\mu,\lambda)}(R,\Gamma|{\NoizeVar})
\notag\\
&=\frac{\lambda(R-\mu \Gamma)
-\underline{\Omega}^{(\mu,\lambda)}({\NoizeVar})}{1+\lambda} 
\MLeq{a}
\frac{\lambda(R-\mu \Gamma)}{1+\lambda} 
\notag\\
& \MLeq{b}
\frac{\lambda (R-\frac{1+\lambda}{2{\NoizeVar}}\Gamma)}{1+\lambda}
 \MEq{c} L^{(\frac{1+\lambda}{2{\NoizeVar}}, \lambda)}(R,\Gamma|{\NoizeVar})
 \MLeq{d} A.
\end{align}
Step (a) follows from that by Lemma \ref{lm:lmSdd},
\begin{align}
& \underline{\Omega}^{(\mu,\lambda)}_{\rmOH}({\NoizeVar})
\geq 
\max_{\eta\geq 0 }\left[ 
\frac{\lambda}{2}\log\left(1+\frac{\eta}{{\NoizeVar}}\right)
\right.
\notag\\
&\qquad\qquad\quad \left. +\frac{1}{2}\log
\left( 1-\frac{\lambda}{1+\lambda}\cdot 2\mu\eta \right)
\right]\geq 0.
\end{align}
Step (b) follows from $\lambda \geq 0$, 
$\mu \geq \frac{1+\lambda}{2{\NoizeVar}}$.
Step (c) follows from the definition of 
$L^{(\mu, \lambda)}(R,\Gamma|{\NoizeVar})$.
Step (d) follows from (\ref{eqn:Asddvv}). 
Thus we have $B\leq A$, completing the proof. 
\hfill\IEEEQED
}
We next drive a relation between $\overline{G}_{\rm AR}
(R,\Gamma|{\NoizeVar})$ and 
$\overline{G}_{\rm DK}(R,\Gamma|{\NoizeVar})$. 
To this end we derive two parametric expressions of 
$\overline{G}_{\rm DK}(R,$ $\Gamma|{\NoizeVar})$. 
For $\mu>0$, we define 
\begin{align}
&\overline{G}_{\rm DK}^{(\mu)}(R,\Gamma|{\NoizeVar}) 
\notag\\
& \defeq
\min_{\scs q_{XY} \in {\cal G}_2}
\big\{
\left[R-I(q_X,q_{Y|X})\right]^+ + D(q_{Y|X}||W|q_X) 
\notag\\ 
& 
\hspace{20mm}
-\mu \left(\Gamma-{\rm E}_{q_X}[X^2] \right)
\big\}. 
\label{objective_function0}
\end{align}
For $\mu, \lambda\geq 0$, we define 
\begin{align}
&\overline{G}_{\rm DK}^{(\mu, \lambda)}(R,\Gamma|{\NoizeVar}) 
\notag\\
& \defeq
\min_{\scs q_{XY} \in {\cal G}_2}
\big\{
\lambda \left[R-I(q_X,q_{Y|X})\right]-\mu\Gamma + \mu {\rm E}_{q_X}[X^2]
\notag\\ 
& 
\hspace{20mm}+ D(q_{Y|X}||W|q_X) \big\}.\label{zzobjective_functionZZ}
\end{align}
According to Oohama \cite{SingleStConvArXiv17}, we have the following lemma.
\begin{lm}[Oohama \cite{SingleStConvArXiv17}] \label{lm:lem2zzz}
For any $R>0$, 
\begin{align}
\overline{G}_{\rm DK}(R,\Gamma|{\NoizeVar}) 
&= \max_{\mu \geq 0} \overline{G}_{\rm DK}^{(\mu)}(R,\Gamma|{\NoizeVar}). 
\label{zzparametric_expressionb}
\end{align}
For any $\mu \geq 0$, any $R>0$, we have
\begin{align}
\overline{G}_{\rm DK}^{(\mu)}(R,\Gamma |{\NoizeVar}) 
&= \max_{0\scs \leq \lambda \leq 1} 
\overline{G}_{\rm DK}^{(\mu,\lambda)}(R,\Gamma|{\NoizeVar}). 
\label{zzparametric_expression2}
\end{align}
The two equalities (\ref{zzparametric_expressionb})
and (\ref{zzparametric_expression2}) imply that 
\begin{align}
\overline{G}_{\rm DK}(R,\Gamma |{\NoizeVar}) 
&= \max_{\scs \mu \geq 0,
\atop{ \scs \lambda \in [0,1]}
}\overline{G}_{\rm DK}^{(\mu,\lambda)}(R,\Gamma|{\NoizeVar}).
\label{zzparametric_expressionc}
\end{align}
\end{lm}

We can show the following proposition 
stating that the two quantities
$\overline{G}_{\rm AR}(R,\Gamma|{\NoizeVar})$ 
and $\overline{G}_{\rm DK}(R,\Gamma|{\NoizeVar})$ 
match. 
\renewcommand{\prmt}{\frac{\lambda}{1+\lambda}}
\newcommand{\barprmt}{\frac{1}{1+\lambda}}
\newcommand{\muprmt}{\frac{\mu\lambda}{1+\lambda}}
\begin{pro}\label{pro:pro1xx}
For any $\mu,\lambda\geq 0$, we have 
the following:
\beqa
\overline{G}_{\rm AR}^{(\mu,\prmt)}(R,\Gamma|{\NoizeVar})
&=&
\overline{G}_{\rm DK}^{(\muprmt,\prmt)}(R,\Gamma|{\NoizeVar}).
\label{eqn:zmainEq00}
\eeqa
In particular, we have 
\beqa
& &
\overline{G}_{\rm AR}(R,\Gamma|{\NoizeVar})
=\overline{G}_{\rm DK}(R,\Gamma|{\NoizeVar}).
\label{eqn:zmainEq0}
\eeqa
\end{pro}

Proof of this proposition is given in Section \ref{sec:ThreeExp}.
\newcommand{\ProofPropositionB}{

{\it Proof of Proposition \ref{pro:pro1xx}: } 
For $q_X \in {\cal G}_1$, we set
$$
{\cal G}_1(q_X)\defeq 
\{ q_{Y|X}:q_{XY}=(q_X,q_{Y|X}) \in {\cal G}_2 \}.
$$

For a given joint Gaussian p.d.f. $q=q_{XY}\in {\cal G}_2$, 
we introduce the conditional p.d.f. 
$q_{X|Y}$ and the p.d.f. $\empty{}q_{Y}$ by 
$$
q_{X}(x)q_{Y|X}(y|x) 
=\empty{}q_{Y}(y)\empty{}q_{X|Y}(x|y).
$$
The above $q_{X|Y}$ is called a backward channel.
Set 
\begin{align*}
& \overline{G}_{\rm DK}^{(\mu\prmt,\prmt)}(R,\Gamma,q_X |{\NoizeVar}) \\
&\defeq  
\min_{\scs q_{Y|X}\in \mathcal{G}_1(q_X)} 
\bigg\{ \prmt (R-\mu\Gamma-I(q_X,q_{Y|X})
\notag\\
&\hspace{2cm}+\mu{\rm E}_{q_X}[X^2]) + D(q_{Y|X}||W| q_X)
\bigg\}. 
\end{align*}
Using $(q_{Y},$ $q_{X|Y})$, 
we obtain the following: 
\beqa
& &\prmt\left\{-I(q_X,q_{Y|X})+\mu{\rm E}_{q_X}[X^2]\right\}
\nonumber\\
& & +D(q_{Y|X}||W|q_X)
\nonumber\\
&=&-\prmt D(q_{X|Y}||q_X |q_{Y})
    +D(\empty{}q_{Y},\empty{}q_{X|Y}||q_X,W) 
\nonumber\\
& &+\muprmt {\rm E}_{(q_{Y},q_{X|Y})}[X^2]
\nonumber\\
&=&\int \int{\rm d}x{\rm d}y q_{Y}(y)q_{X|Y}(x|y)
\log\left\{\frac{\empty{}q_{X|Y}^{-\prmt}(x|y)}{q_X^{-\prmt}(x)} 
    \right\}
\nonumber\\
& & +\int \int{\rm d}x{\rm d}y q_{Y}(y)q_{X|Y}(x|y)
\nonumber\\
& &\qquad\times
    \log \left\{ 
     \frac{\empty{}q_{X|Y}(x|y) q_{Y}(y)}
     {q_X(x)W(y|x){\rm e}^{-\muprmt x^2}} 
     \right\}
\nonumber\\
\newcommand{\SDff}
{
&=&\int\int{\rm d}x{\rm d}y q_{Y}(y)q_{X|Y}(x|y)
\nonumber\\
& &\qquad\times
\log\left\{
\frac{\empty{}q_{X|Y}^{\barprmt}(x|y)}
     {q_X^{\barprmt}(x)W(y|x){\rm e}^{-\muprmt x^2}} 
    \right\}
\nonumber\\
& &+\int{\rm d}y 
q_{Y}(y) \log q_{Y}(y)
\nonumber\\
}
&=&\barprmt \int \int{\rm d}x{\rm d}y q_{Y}(y)q_{X|Y}(x|y)
\nonumber\\
& &\qquad\times
    \log\left\{
\frac{q_{X|Y}(x|y)}
{q_X(x)\left\{W(y|x){\rm e}^{-\muprmt x^2} \right \}^{1+\lambda}} 
        \right\}
\nonumber\\
& &+\int{\rm d}y q_{Y}(y) \log \empty{}q_{Y}(y)
\nonumber\\
&=&\barprmt D(\empty{}q_{X|Y}||\hat{q}_{X|Y}|q_{Y})
    +D(\empty{}q_{Y}||\hat{q}_{Y}) 
\nonumber\\
& &
- J^{(\mu,\prmt)}(q_X \Vl {\NoizeVar}), 
\label{eqn:ZZequal0}
\eeqa
where 
$\hat{q}_{X|Y}$ $=\hat{q}_{X|Y}(x|y)$ 
is a conditional p.d.f. given by 
\beqa
\hat{q}_{X|Y}(x|y)
&=&\frac{1}{\Lambda(y)}q_{X}(x){
\left\{W(y|x){\rm e}^{-\muprmt x^2}\right\}
}^{1+\lambda},
\label{eqn:ZZdist0}\\
\Lambda(y) &\defeq & \int {\rm d}x q_X(x)
\left\{W(y|x){\rm e}^{-\muprmt x^2}\right\}^{1+\lambda},
\nonumber
\eeqa
and $\hat{q}_Y$ $=\left\{\hat{q}_Y(y)\right\}_{y\in{\cal Y}}$ 
is a p.d.f. given by
\beq
\hat{q}_Y(y)=\frac{\Lambda(y)^{\barprmt}}
         { \int{\rm d}y \Lambda(y)^{\barprmt}}.
\label{eqn:ZZdist1}
\eeq
Hence, by (\ref{eqn:ZZequal0}) and the non-negativity of divergence, 
we obtain 
$$
\overline{G}_{\rm DK}^{(\muprmt,\prmt)}(R,\Gamma, q_X \Vl \NoizeVar)\geq 
{\empty}{G}_{\rm AR}^{(\mu,\prmt)}(R,\Gamma, q_X \Vl \NoizeVar)
$$ 
for any $q_X \in {\cal G}_1$. 
Hence we have 
$$
 \overline{G}_{\rm DK}^{(\muprmt,\prmt)}(R,\Gamma \Vl \NoizeVar)
\geq \overline{G}_{\rm AR}^{(\mu,\prmt)}(R,\Gamma \Vl \NoizeVar).
$$ 
We next prove that for any $\lambda \geq 0$,
$$
\overline{G}_{\rm DK}^{(\muprmt,\prmt)}(R,\Gamma \Vl W) 
\leq \overline{G}_{\rm AR}^{(\mu, \prmt)}( R,\Gamma \Vl W).
$$
Let $q_{X,\theta}$ be a Gaussian distribution 
with mean 0 and variance $\theta$. 
We assume that $q_{X,\theta}$ attains 
$\overline{G}_{\rm AR}^{(\mu,\prmt)}(R,\Gamma \Vl W)$.
Then by Proposition \ref{pro:proDss}, we have 
\beqa
\xi&=&\xi(\mu,\lambda, \theta)
= \frac{(1+\lambda)\theta}{1+2\mu\lambda\theta}
=\frac{1}{2\mu}-\frac{{\NoizeVar}}{1+\lambda}.
\label{eqn:SzzRR}
\eeqa
From (\ref{eqn:SzzRR}), we have 
\beq
\mu=\frac{1}{2}\frac{1+\lambda}{(1+\lambda)\xi+\NoizeVar},
\theta=\frac{\xi}{(1+\lambda)-2\mu\lambda \xi}.
\label{eqn:asEddx}
\eeq
From 
(\ref{eqn:asEddx}), we have
\beq
\theta=\frac{[(1+\lambda)\xi+\NoizeVar]\xi}
{(1+\lambda)(\xi+\NoizeVar)}.
\label{eqn:asEddxx}
\eeq
Computing $\Lambda(y)$ for this $q_{X,\theta}$, we have
\begin{align}
&\Lambda(y)=\int {\rm d}x q_{X,\theta}W^{1+\lambda}(y|x){\rm e}^{-\mu \lambda x^2}
\notag\\
&=\frac{1}{\sqrt{2\pi\theta}}
\left(\frac{1}{\sqrt{2\pi\NoizeVar}}\right)^{1+\lambda}
\int{\rm d}x
\notag\\
& \quad \times 
{\rm e}^{
-\frac{1}{2}\left[
\frac{x^2}{\theta}+(1+\lambda)\frac{(x-y)^2}{\NoizeVar}+2\mu\lambda x^2
\right]}
\notag\\
&\MEq{a}\frac{1}{\sqrt{2\pi\theta}}
\left(\frac{1}{\sqrt{2\pi\NoizeVar}}\right)^{1+\lambda}
\int{\rm d}x
\notag\\
& \quad \times 
{\rm e}^{
-\frac{1+\lambda}{2}
\left[\frac{\xi+\NoizeVar}{\xi\NoizeVar}
\left(x-\frac{\xi}{\xi+\NoizeVar}y\right)^2
+\frac{y^2}{\xi+\NoizeVar}\right]}
\notag\\
&=
\left(\frac{1}{\sqrt{2\pi\NoizeVar}}\right)^{1+\lambda}
\sqrt{\frac{1}{\theta(1+\lambda)}\frac{\xi\NoizeVar}{\xi+\NoizeVar}}
{\rm e}^{-\frac{1+\lambda}{2}\cdot\frac{y^2}{\xi+{\NoizeVar}}}
\notag\\
&\MEq{b}
\left(\frac{1}{\sqrt{2\pi\NoizeVar}}\right)^{1+\lambda}
\sqrt{\frac{\NoizeVar}{(1+\lambda)\xi+\NoizeVar}}
{\rm e}^{-\frac{1+\lambda}{2}\cdot\frac{y^2}{\xi+{\NoizeVar}}}.
\label{eqn:AssDz}
\end{align}
Step (a) follows from (\ref{eqn:Asssjj}). Step (b) follows 
from (\ref{eqn:asEddxx}). Using (\ref{eqn:AssDz}), we have 
the following: 
\begin{align}
& \int {\rm d}y \Lambda(y)^{\barprmt}
= \left[
\frac{\NoizeVar}{(1+\lambda)\xi+\NoizeVar}
\right]^{\frac{1}{2(1+\lambda)}}
 \notag\\
&\quad \times 
\frac{1}{\sqrt{2\pi\NoizeVar}}
\int {\rm d}y 
{\rm e}^{-\frac{1+\lambda}{2}\cdot\frac{y^2}{\xi+{\NoizeVar}}}
\notag\\
&=
\left[
\frac{\NoizeVar}{(1+\lambda)\xi+\NoizeVar}
\right]^{\frac{1}{2(1+\lambda)}}
\sqrt{\frac{\xi+\NoizeVar}{\NoizeVar}}.
\label{eqn:Saatt}
\end{align}
Using (\ref{eqn:AssDz}), we also have 
\begin{align}
&\int {\rm d}y \left\{
W(y|x){\rm e}^{-\muprmt x^2}
\right\}^{1+\lambda}\Lambda(y)^{-\frac{\lambda}{1+\lambda}}
\notag\\
&=\left[
\frac{\NoizeVar}{(1+\lambda)\xi+\NoizeVar}
\right]^{\frac{-\lambda}{2(1+\lambda)}}
\left(\frac{1}{\sqrt{2\pi\NoizeVar}}\right)^{1+\lambda}
\notag\\
&\quad \times \left({\sqrt{2\pi\NoizeVar}}\right)^{\lambda}\int{\rm d}y
{\rm e}^{
-\frac{1}{2}\frac{1+\lambda}{\NoizeVar}(y-x)^2
+\frac{\lambda}{2(\xi+\NoizeVar)} y^2-\mu \lambda x^2}
\notag\\
&=\left[
\frac{\NoizeVar}{(1+\lambda)\xi+\NoizeVar}
\right]^{\frac{-\lambda}{2(1+\lambda)}}
\frac{1}{\sqrt{2\pi\NoizeVar}}
\notag\\
&\quad \times \int{\rm d}y
{\rm e}^{
-\frac{1}{2}\frac{(1+\lambda)\xi+\NoizeVar}{\NoizeVar(\xi+\NoizeVar)}
\left(y-\frac{(1+ \lambda)(\xi+\NoizeVar)}{(1+\lambda)\xi+\NoizeVar}x\right)^2
}
\notag\\
&\qquad \times {\rm e}^{
 \frac{1}{2}\frac{\lambda(1+\lambda)}{(1+\lambda)\xi+\NoizeVar}x^2
-\mu\lambda x^2
}
\notag\\
&\MEq{a}\left[
\frac{\NoizeVar}{(1+\lambda)\xi+\NoizeVar}
\right]^{\frac{-\lambda}{2(1+\lambda)}}
\frac{1}{\sqrt{2\pi\NoizeVar}}
\notag\\
&\quad \times \int{\rm d}y
{\rm e}^{
-\frac{1}{2}
\frac{(1+\lambda)\xi+\NoizeVar}{\NoizeVar(\xi+\NoizeVar)}
\left(y-\frac{(1+ \lambda)(\xi+\NoizeVar)}{(1+\lambda)\xi+\NoizeVar}x\right)^2
}
\notag\\
&=
\left[
\frac{\NoizeVar}{(1+\lambda)\xi+\NoizeVar}
\right]^{\frac{1}{2(1+\lambda)}}
\sqrt{\frac{\xi+\NoizeVar}{\NoizeVar}}.
\label{eqn:ZZcond0}
\end{align}
Step (a) follows from the first equality of (\ref{eqn:asEddx}). 
From (\ref{eqn:Saatt}) and (\ref{eqn:ZZcond0}), we have 
\beq
\int {\rm d}y \left\{
W(y|x){\rm e}^{-\muprmt x^2}
\right\}^{1+\lambda}\Lambda(y)^{-\frac{\lambda}{1+\lambda}}
=
\int {\rm d}y \Lambda(y)^{\barprmt}.
\label{eqn:qsddA}
\eeq
Now we define the function ${V}=V(y|x)$ 
by
\beq
V(y|x)=\frac{\hat{q}_Y(y)\hat{q}_{X|Y}(x|y)}{q_X(x)}.
\label{eqn:ZZvdist}
\eeq
By (\ref{eqn:ZZdist0}) and (\ref{eqn:ZZdist1}), 
$V(y|x)$ has the following form:
\begin{align}
& V(y|x)
=\frac{\Lambda(y)^{\barprmt}}
   {\int{\rm d}y  \Lambda(y)^{\barprmt}}
\nonumber\\
& \qquad \times
   \frac{1}{\Lambda(y)}q_{X}(x)
\left\{W(y|x){\rm e}^{-\muprmt x^2}\right\}^{1+\lambda}
   \cdot \frac{1}{q_X(x)}
\nonumber\\
&= \frac{
\left\{W(y|x){\rm e}^{-\muprmt x^2}\right\}^{1+\lambda}
         \Lambda(y)^{-\prmt}}
        {\int{\rm d}y \Lambda(y)^{\barprmt}}.
\label{eqn:ZZdist4}
\end{align}
Taking integral of both sides of (\ref{eqn:ZZdist4}) 
on the variable $y$, we obtain
\beqno
& &\int{\rm d}y V(y|x)
\\
&=& \frac{\int{\rm d}y 
\left\{ W(y|x) {\rm e}^{-\muprmt x^2}
\right\}^{1+\lambda} \Lambda(y)^{-\prmt}}
   { \int{\rm d}y \Lambda(y)^{\barprmt} }
\MEq{a}1.  
\eeqno
Step (a) follows from (\ref{eqn:qsddA}). The above equality 
implies that $V$ is a conditional 
density function. Furthermore, note that from (\ref{eqn:ZZvdist}),
$$
q_{X}(x)V(y|x)=\hat{q}_{Y}(y)\hat{q}_{X|Y}(x|y).
$$
Then, choosing $q_{Y}=\hat{q}_Y,q_{X|Y}=\hat{q}_{X|Y}$ 
in (\ref{eqn:ZZequal0}), we have, for $\lambda \geq 0$, 
\beqno
& &\overline{G}_{\rm DK}^{(\mu,\prmt)}(R, \Gamma \Vl W) 
\\
&\leq &\prmt\left\{(R-\mu \Gamma) -I(q_X,V)+ \mu{\rm E}_{q_X} [c(X)] \right\}
\\
& &+ D(V||W|q_X)
\nonumber\\ 
&=&\prmt(R-\mu \Gamma) - J^{(\mu, \prmt)}(q_X \Vl W)
=\overline{G}_{\rm AR}^{(\mu, \prmt)}(R,\Gamma \Vl W), 
\eeqno
completing the proof. \hfill\IEEEQED 
}
From Theorems \ref{Th:GauDK}, \ref{Th:GauMain} and  
Propositions \ref{pro:proDss}, \ref{pro:pro1xx}, we immediately obtain the 
following theorem. 
\begin{Th}
\beqno
& &G^*(R,\Gamma|{\NoizeVar}) 
\nonumber\\
&=& G_{\rmOH}(R,\Gamma|{\NoizeVar})
   =G_{\rm AR}(R,\Gamma|{\NoizeVar})
   =\overline{G}_{\rm DK}(R,\Gamma|{\NoizeVar})
\nonumber\\
&=&\max_{\scs \lambda \geq 0,
\atop{\scs
\mu \in [0, \frac{1+\lambda}{2{\NoizeVar}}]}}
L^{(\mu,\lambda)}(R,\Gamma|{\NoizeVar}).
\eeqno
\end{Th}

We finally solve the maximization problem defining 
$G($ $R,\Gamma|{\NoizeVar})$ to derive an explicit 
formula of this exponent function. 
We set
\beq
\ba{l}
\ds \rho \defeq \frac{\lambda}{1+\lambda},
\ds \nu \defeq \frac{2\lambda \mu {\NoizeVar}}
{1+\lambda}=\rho\cdot(2\mu {\NoizeVar}).
\ea
\eeq
Then we have 
\beqa
& &\lambda \geq 0, \mu 
\in \left[0,\frac{1+\lambda}{2\NoizeVar}\right]  
\Leftrightarrow 0\leq \frac{\nu}{1+\nu}\leq \rho < 1, 
\label{eqn:Cond1}\\
& &L^{(\mu,\lambda)}(R,\Gamma|{\NoizeVar})
=\rho R -\frac{\nu}{2} \cdot \frac{\Gamma}{{\NoizeVar}}
-\frac{1}{2}\log(1+\nu)
\nonumber\\
& &\qquad
   + \frac{\rho}{2}\log \nu +\frac{1}{2}h(\rho), 
\label{eqn:ObFunc1}
\eeqa
where $h(\cdot)$ is the binary entropy function.
Since by (\ref{eqn:ObFunc1}), we can regard 
$L^{(\mu,\lambda)}(R,\Gamma|{\NoizeVar})$ as a quantity with 
parametor $\rho$ and $\nu$, we denote 
it by $L^{(\rho,\nu)}(R,\Gamma|{\NoizeVar})$.
Then we have 
\beq
G(R,\Gamma |\NoizeVar)
=\max_{\rho\in[ \frac{\nu}{1+\nu},1)}
L^{(\rho,\nu)}(R,\Gamma|{\NoizeVar}). 
\label{eqn:Xddf}
\eeq
Solving the optimization problem (\ref{eqn:Xddf})
defining $G(R,\Gamma|\NoizeVar)$, we obtain the 
following result.

\begin{Th}\label{th:PrmtTh}
Let $\nu_0$ be the unique positive solution of 
$$
\nu_0(1+\nu_0)=\frac{\sigma^2}{\Gamma}.
$$
Then, the exponent function $G(R,\Gamma|\NoizeVar)$ has 
the following parametric form:
\beq
\left.
\ba{rcl}
R&=&\ds \frac{1}{2}\log 
\frac{\ds 1+\frac{\Gamma}{\NoizeVar}(1+\nu)}
{\ds1-\frac{\Gamma}{\NoizeVar}\nu(1+\nu)},
\vspace*{1mm}\\
G(R,\Gamma|\NoizeVar)
&=&\ds -\frac{\nu \Gamma}{2\NoizeVar}-\frac{1}{2}
\log\left[1-\frac{\Gamma}{\NoizeVar}\nu(1+\nu)\right],
\vspace{2mm}\\
& & \mbox{ for some }\nu\in [0,\nu_0).
\ea
\right\}
\eeq
\end{Th}

Proof of this theorem is given in Section \ref{sec:ThreeExp}.
\newcommand{\PrfofPrmF}{

{\it Proof of Theorem \ref{th:PrmtTh}:}  
We set 
\begin{align}
& F=F(\rho,\nu)=-L^{(\rho,\nu)}(R,\Gamma|{\NoizeVar}) 
\notag\\
&=-\rho R +\frac{\nu}{2} \cdot \frac{\Gamma}{{\NoizeVar}}
  +\frac{1}{2}\log(1+\nu)
- \frac{\rho}{2}\log \nu-\frac{1}{2}h(\rho).
\end{align}
Then we have 
\beqno
G(R,\Gamma |\NoizeVar)
=(-1)\cdot\min_{\rho\in [\frac{\nu}{1+\nu},1) }
F(\rho,\nu).
\eeqno
Computing the gradient of $F$, we obtain
\beq
\left.
\ba{rcl}
& &\ds \frac{\partial F}{\partial \rho}=-R-\frac{1}{2}\log \nu 
+\frac{1}{2}\log \frac{\rho}{1-\rho},
\vspace{1mm}\\
& &\ds \frac{\partial F}{\partial \nu}= 
\frac{1}{2}\left[
\frac{\Gamma}{\NoizeVar}+\frac{1}{1+\nu}-\frac{\rho}{\nu}
\right].
\ea
\right\}
\label{eqn:Xddfa}
\eeq
Let 
$$
B=\left( 
\ba{cc}
\ds \frac{\partial^2 F}{\partial \rho^2}
& \ds \frac{\partial^2 F}{\partial \rho \partial \nu}
\vspace{1mm}\\
\ds \frac{\partial^2 F}{\partial \rho \partial \nu}
& \ds \frac{\partial^2 F}{\partial \nu^2}
\ea
\right)
$$ 
be the Hessian matrix of $F$. Computing $B$, we have 
\beqa
& & \frac{\partial^2 F}{\partial \rho^2}
=\frac{1}{2\rho(1-\rho)}>0,
\label{eqn:AssPa}\\
& & \frac{\partial^2 F}{\partial \rho \partial \nu}=-\frac{1}{2\nu},
\nonumber\\
& & \frac{\partial^2 F}{\partial \nu^2}=
\frac{1}{2}\left[
-\frac{1}{(1+\nu)^2}
+\frac{\rho}{\nu^2}
\right]
\nonumber\\
&&\MGeq{a} 
\frac{1}{2}\left[-
\frac{1}{(1+\nu)^2}+\frac{1}{(1+\nu)\nu}
\right]=\frac{1}{2(1+\nu)^2\nu}>0.\quad 
\label{eqn:AssPb}
\eeqa
Step (a) follows from $\rho\in [\nu/(1+\nu),1)$. 
Computing $|B|$, we have 
\beqa
& &|B|=
\ds \frac{\partial^2 F}{\partial \rho^2}
\frac{\partial^2 F}{\partial \nu^2}
-\left(\frac{\partial^2 F}{\partial \rho \partial \nu}\right)^2
\nonumber\\ 
&=&\frac{1}{\rho(1-\rho)}\left[
\frac{\rho}{\nu^2}-\frac{1}{(1+\nu)^2}\right]
-\frac{1}{\nu^2}
\nonumber\\
&=&\frac{1}{\rho(1-\rho)}
\left[\frac{\rho^2}{\nu^2}-\frac{1}{(1+\nu)^2}\right]
\nonumber\\
&=&\frac{1}{\rho(1-\rho)}
\left[\frac{\rho}{\nu}+\frac{1}{1+\nu}\right]
\left[\frac{\rho}{\nu}-\frac{1}{1+\nu}\right]
\MGeq{a}0.
\label{eqn:AssPc}
\eeqa
Step (a) follows from $\rho\in [\nu/(1+\nu),1)$. From (\ref{eqn:AssPa}), 
(\ref{eqn:AssPb}), and (\ref{eqn:AssPc}), we can see that $F(\rho,\nu)$ is 
a convex function of $(\rho,\nu)$. Hence from (\ref{eqn:Xddfa}), we have 
that if the pair $(\rho^*,\nu^*)$ satisfies 
$\rho^*\in [\nu^*/(1+\nu^*),1)$ and 
\beq
\left.
\ba{rcl}
& &\ds \left. 
\frac{\partial F}{\partial \rho}\right|_{(\rho,\nu)=(\rho^*,\nu^*)}= 
-R-\frac{1}{2}\log \nu^* 
+\frac{1}{2}\log \frac{\rho^*}{1-\rho^*}=0,
\vspace{1mm}\\
& &\ds \left. \frac{\partial F}{\partial \nu}\right|_{
(\rho,\nu)=(\rho^*,\nu^*)}=  
\frac{1}{2}\left[
\frac{\Gamma}{\NoizeVar}+\frac{1}{1+\nu^*}-\frac{\rho^*}{\nu^*}
\right]=0,
\ea
\right\}
\label{eqn:Xddfaz}
\eeq
then, it attains the minimum of $F(\rho,\nu)$ under 
$\rho\in [\nu/(1+\nu),1)$. 
From (\ref{eqn:Xddfaz}), we have 
\beqa
R&=&-\frac{1}{2}\log \nu^* 
+\frac{1}{2}\log \frac{\rho^*}{1-\rho^*},
\label{eqn:zsfaz2}
\\
\rho^* &=&\frac{\nu^*}{1+\nu^*}+ \nu^*\frac{\Gamma}{\NoizeVar}
\geq \frac{\nu^*}{1+\nu^*}.
\label{eqn:zsfaz}
\eeqa
Furthermore, for $\nu^*\in [0,\nu_0)$, we have 
\begin{align}
& 1-\rho^*=\frac{1}{1+\nu^*}-\nu^*\frac{\Gamma}{\NoizeVar}
\notag\\
&=\frac{\ds 1-\nu^*(1+\nu^*)\frac{\Gamma}{\NoizeVar}}{1+\nu^*}
>\frac{\ds 1-\nu_0(1+\nu_0)\frac{\Gamma}{\NoizeVar}}{1+\nu^*}=0.
\label{eqn:zsfaz3}
\end{align}
From (\ref{eqn:zsfaz2})-(\ref{eqn:zsfaz3}), we can see that for 
$\nu^*\in [0,\nu_0)$, the pair $(\rho^*,\nu^*)$ certainly attains 
$G(R,\Gamma|\NoizeVar)=F(\rho^*,\nu^*)$. From (\ref{eqn:zsfaz2}) 
and (\ref{eqn:zsfaz}), we have 
\beq
R=\frac{1}{2}\log 
\frac{\ds 1+\frac{\Gamma}{\NoizeVar}(1+\nu^*)}
{\ds1-\frac{\Gamma}{\NoizeVar}\nu^*(1+\nu^*)}.
\label{eqn:ZsdP}
\eeq
Furthermore, we have
\beqa
& &G(R,\Gamma|\NoizeVar)=-F(\rho^*,\nu^*)
\nonumber\\
&=&\rho^* R-\frac{\nu^* \Gamma}{2\NoizeVar}
- \frac{1}{2} \log (1+\nu^*)
+\frac{1}{2} \rho^* \log \nu^* 
+\frac{1}{2}h(\rho^*)
\nonumber\\
&=&\rho^*\left[R+\frac{1}{2}\log \nu^* 
+\frac{1}{2}\log \frac{1-\rho^*}{\rho^*}\right]
\nonumber\\
& &-\frac{\nu^* \Gamma}{2\NoizeVar}-\frac{1}{2}\log(1+\nu^*)
-\frac{1}{2}\log(1-\rho^*)
\nonumber\\
&\MEq{a}&-\frac{\nu^* \Gamma}{2\NoizeVar}-\frac{1}{2}
\log\left[1-\frac{\Gamma}{\NoizeVar}\nu^*(1+\nu^*)\right].
\label{eqn:ZsdPb}
\eeqa
Step (a) follows from (\ref{eqn:zsfaz2}) and (\ref{eqn:zsfaz}).
Thus we obtain the parametric expression of $G(R,\Gamma|\NoizeVar)$
given by (\ref{eqn:ZsdP}) and (\ref{eqn:ZsdPb}).
\hfill\IEEEQED
}
\newcommand{\ApdaAAC}{}
\newcommand{\DelTh}{
aaaa
}
\section{Proofs of Results}
\label{sec:ThreeExp}

In this section we prove Propositions 
\ref{pro:proDss}, \ref{pro:pro1xx} and Theorem \ref{th:PrmtTh} 
stated in Section \ref{sec:MainResult}.
\LemmaForProposition
\ProofPropositionA
\ProofPropositionB
\PrfofPrmF

\section*{\empty}

\appendix


\ApdaDirectDK


\newcommand{\ZapXXXXXXXX}{
\section{Equivalence of Three Exponent Functions} 

In this section we prove Lemmas \ref{lm:lmSddQ}, \ref{lm:lem2}, and 
Proposition \ref{pro:pro1} stated in Section II.  
We first prove Lemma \ref{lm:lmSddQ}.

{\it Proof of Lemma \ref{lm:lmSddQ}: } We observe that
\beqa
\Omega^{(\mu,\lambda)}(W)
&=&\max_{q_X \in{\cal P}({\cal X})}
\log \biggl\{ \min_{Q \in{\cal P}({\cal Y})}
\sum_{x,y} q_{X}(x)W(y|x)
\nonumber\\
& &\qquad \times
\left[\frac{W(y|x){\rm e}^{-\mu c(x)}}{Q(y)} \right]^\lambda
\biggr\}
\label{eqn:aAzzb}. 
\eeqa
On the objective function of the minimization problem 
inside the logarithm function in (\ref{eqn:aAzzb}), 
we have the following 
chain of inequalities:
\beqa
&&\sum_{x,y}q_{X}(x)W(y|x)
\left[\frac{W(y|x){\rm e}^{-\mu c(x)}}{Q(y)} \right]^\lambda
\nonumber\\
&=&\sum_{y}\left[
\sum_{x}q_X(x)W^{1+\lambda}(y|x){\rm e}^{-\muprmt c(x)}\right]Q^{-\lambda}(y)
\nonumber\\
&\MGeq{a}&
\left\{\sum_{y}
\left[\sum_{x}q_X(x) W^{1+\lambda}(y|x)
{\rm e}^{-\muprmt c(x)}
\right]^{\frac{1}{1+\lambda}}
\right\}^{1+\lambda} 
\nonumber\\
&&\quad \times
\left\{ \sum_{y}Q(y)\right\}^{-\lambda}
\nonumber\\
&=&
\left\{\sum_{y}
\left[
\sum_{x}q_X(x) 
W^{1+\lambda}(y|x){\rm e}^{-\muprmt c(x)}\right]^{\frac{1}{1+\lambda}}
\right\}^{1+\lambda}
\nonumber\\
&=&
\exp\left\{(1+\lambda)J^{(\mu,\frac{\lambda}{1+\lambda})}(q_X|{\NoizeVar}) \right\}.
\label{eqn:AzzzV}
\eeqa    
In (a), we have used the reverse H\"older inequality
$$
\sum_i a_ib_i\geq 
\left( \sum_{i} a_i^{\frac{1}{\alpha}}\right)^{\alpha}
\left(\sum_{i} b_i^{\frac{1}{\beta}}\right)^{\beta}
$$
which holds for nonegative $a_i,b_i$ and for $\alpha+\beta=1$ 
such that either $\alpha >1$ or $\beta >1$. In our case we 
have applied the inequality to 
$$
\left.
\ba{rcl}
i &\to& y,
\vspace*{1mm}\\
a_i &\to& \ds 
\sum_{x}q_X(x) 
W^{1+\lambda}(y|x){\rm e}^{-\muprmt c(x)},
\vspace*{1mm}\\
b_i &\to& Q^{-\lambda}(y), 
\vspace*{1mm}\\
(\alpha,\beta) &\to& (1+\lambda, -\lambda).
\ea
\right\}
$$
In the reverse H\"older inequality the equality holds if and only if 
$a_i^{\frac{1}{\alpha}}=\kappa b_i^{\frac{1}{\beta}}$ for some constant 
$\kappa$. In (\ref{eqn:AzzzV}), the equality holds for 
$$
Q(y)=\kappa \left[ 
\sum_{x}q_X(x) 
W^{1+\lambda}(y|x){\rm e}^{-\muprmt c(x)}\right]^{\frac{1}{1+\lambda}},
$$
where $\kappa$ is a normalized constant. 
From (\ref{eqn:AzzzV}), we have
\beqno
& &
\Omega^{(\mu,\lambda)}(W)
=(1+\lambda)
\max_{q_X \in{\cal P}({\cal X})}J^{(\mu,\frac{\lambda}{1+\lambda})}
(q_X|{\NoizeVar}),
\eeqno
completing the proof.
\hfill\IEEEQED

We can show that $G_{\rm DK}(R,\Gamma|{\NoizeVar})$ satisfies 
the following property. 
\begin{pr}\label{pr:prOne}$\quad$ 
\begin{itemize}
\item[\rm{a)}] 
For every fixed $\Gamma>0$, the function ${G}_{\rm DK}(R,\Gamma \Vl W)$ 
is monotone increasing for $R \geq 0$ and takes positive value 
if and only if $R > C(\Gamma | W)$. For every fixed $R \geq 0$, the 
function ${G}_{\rm DK}(R,\Gamma \Vl W)$ is monotone decreasing 
for $\Gamma >0.$

\item[{\rm b)}] 
${G}_{\rm DK}(R, \Gamma \Vl W)$ is a convex function of $(R,\Gamma)$.

\item[{\rm c)}] For $ R, R^{\prime} \geq 0$
$$
| {G}_{\rm DK }(R, \Gamma \Vl W)-{G}_{\rm DK}(R^{\prime}, \Gamma \Vl W)|
\leq |R-R^{\prime}|.
$$
\end{itemize}
\end{pr}

Property \ref{pr:prOne} part a) is obvious. 
See Appendix for the proof of the part b).
Proof of part c) is quite similar to that of 
the case without input cost given by Dueck and 
K\"orner \cite{Dueck_Korner1979}. We omit the detail.

We can show that $G_{\rm DK}^{(\mu)}(R,\Gamma|{\NoizeVar})$ satisfies 
the following property.
\begin{pr}\label{pr:prTwoZZ} $\quad$ 
\begin{itemize}
\item[\rm{a)}] 
For every fixed $\Gamma>0$, the function 
${G}_{\rm DK}^{(\mu)}(R,\Gamma \Vl W)$ is monotone 
increasing for $R \geq 0 $.
For every fixed $R\geq 0$, the function 
${G}_{\rm DK}^{(\mu)}(R,\Gamma \Vl W)$ is monotone 
decreasing for $\Gamma >0.$
\item[{\rm b)}] For every fixed $\mu\geq 0$, the function 
${G}_{\rm DK}^{(\mu)}(R, \Gamma \Vl W)$ is a convex 
function of $(R,\Gamma)$.
\item[{\rm c)}] For $ R, R^{\prime} \geq 0$
$$
| {G}_{\rm DK }^{(\mu)}(R, \Gamma \Vl W)-{G}^{(\mu)}_{\rm DK}
(R^{\prime}, \Gamma \Vl W)|
\leq |R-R^{\prime}|.
$$
\end{itemize}
\end{pr}

Property \ref{pr:prTwoZZ} part a) is obvious. 
See Appendix for the proof of the part b).
Proof of part c) is quite similar to that of 
the case without input cost given by Dueck and 
K\"orner \cite{Dueck_Korner1979}. We omit the detail.

{\it Proof of (\ref{parametric_expressionb}) in Lemma \ref{lm:lem2}: }
From its formula, it is obvious that for any $\mu\geq 0$
$$
{G}_{\rm DK}(R,\Gamma \Vl W) 
\geq 
{G}_{\rm DK}^{(\mu)}(R,\Gamma \Vl W).
$$
Hence it suffices to prove that for 
any $\Gamma > 0$, there exists $\mu \geq 0$ such that
\begin{equation}
{G}_{\rm DK} (R,\Gamma \Vl W) 
\leq {G}_{\rm DK}^{(\mu)}(R,\Gamma \Vl W).
\label{Gsp=GdeltazzB}
\end{equation}
By Property \ref{pr:prOne} part b), ${G}_{\rm DK}(R, \Gamma \Vl W)$ 
is a monotone decreasing and convex function 
of $\Gamma$. 
Then, there exists $\mu \geq 0$ such that for any 
$\Gamma^{\prime}$ $\geq 0$, we have 
\begin{equation}
{G}_{\rm DK}(R,\Gamma^{\prime} \Vl W) 
\geq {G}_{\rm DK}(R,\Gamma \Vl W)
-\mu(\Gamma^{\prime}-\Gamma). 
\label{ineq_GspB}
\end{equation}
Fix the above $\mu$. 
Let $q^* \in {\cal P}({\cal X}\times{\cal Y})$
be a joint distribution that attains 
$G_{\rm DK}^{(\mu)}(R,\Gamma \Vl W)$.
Set $\Gamma^{\prime}={\rm E}_{q^*}[c(X)]$. 
By the definition of ${G}_{\rm DK}(R,\Gamma' \Vl W)$,
we have
\beqa
&& {G}_{\rm DK}(R,\Gamma' \Vl W)
\nonumber\\
&\leq&
\left[R-I(q^*_X,q^*_{Y|X})\right]^+ + D(q^*_{Y|X}||W|q^*_X). 
\label{ByDefOfGspB}  
\eeqa
Then, we the following chain of inequalities: 
\begin{align}
& G_{\rm DK}(R,\Gamma|{\NoizeVar}) 
\stackrel{\rm (a)}{\leq} 
G_{\rm DK}(R,\Gamma' |{\NoizeVar})+\mu(\Gamma'-\Gamma) 
\notag\\
\stackrel{\rm (b)}{\leq} & 
[R- I({q_{X}^*}, {q_{Y|X}^*})]^+ +D({q^*_{Y|X}} || W | {q_X^*})  
+\mu( \Gamma'-\Gamma) 
\notag\\
\stackrel{\rm (c)}{=} & 
[R- I({q_{X}^*}, {q_{Y|X}^*})]^+ +D({q^*_{Y|X}} || W | {q_X^*})  
\notag\\
 & -\mu(\Gamma - {\rm E}_{q^*}[c(X)]) 
= G_{\rm DK}^{(\mu)}(R,\Gamma \Vl W).
\label{ineq_GlambdaB}
\end{align}
Step (a) follows from (\ref{ineq_GspB}).
Step (b) follows from (\ref{ByDefOfGspB}). 
Step (c) follows from the choice of $\Gamma'={\rm E}_{q^*}[c(X)]$. 
It follows from (\ref{ineq_GlambdaB}) that 
for $\Gamma > 0$,
(\ref{Gsp=GdeltazzB}) holds for some $\mu\geq 0$.
This completes the proof. 
$\quad$\hfill\IEEEQED

{\it Proof of (\ref{parametric_expression2}) in Lemma \ref{lm:lem2}: }
Since $[a]^{+}\geq \lambda a$ for any $a$ and any $\lambda \in [0,1]$, 
it is obvious that 
$$
{G}_{\rm DK}^{(\mu)}(R,\Gamma \Vl W) 
\geq 
\max_{0 \leq  \lambda \leq 1} {G}_{\rm DK}^{(\mu,\lambda)}(R,\Gamma \Vl W).
$$
Hence it suffices to prove that for $R\geq 0$, there exists 
$\lambda \in [0,1]$ such that
$
{G}_{\rm DK}^{(\mu)} (R,\Gamma \Vl W) 
\leq {G}_{\rm DK}^{(\mu,\lambda)}(R,\Gamma \Vl W).
$
By Property \ref{pr:prTwoZZ} part b) 
${G}_{\rm DK}^{(\mu)}(R, \Gamma \Vl W)$ 
is a monotone increasing and convex function of $R$. 
Then, by Property \ref{pr:prOne} part c), 
there exists $0 \leq  \lambda \leq 1$ such that for any 
$R^{\prime}$ $\geq 0$, we have 
\begin{equation}
{G}_{\rm DK}^{(\mu)}(R^{\prime},\Gamma \Vl W) 
\geq {G}_{\rm DK}^{(\mu)}(R,\Gamma \Vl W)
+ \lambda(R^{\prime}-R). \label{ineq_Gsp}
\end{equation}
Let $q^* \in {\cal P}({\cal X}\times{\cal Y})$
be a joint distribution that attains 
$G_{\rm DK}^{(\mu, \prmt)}($ $R,\Gamma \Vl W)$.
Set $R^{\prime}=I({q_X^*}, {q_{Y|X}^*})$. 
Then we have the following chain of inequalities: 
\begin{align}
& G_{\rm DK}^{(\mu)}(R,\Gamma |{\NoizeVar}) 
\stackrel{\rm (a)}{\leq} 
G_{\rm DK}^{(\mu)}(R',\Gamma |{\NoizeVar}) - \lambda(R'-R) 
\notag\\
= & \min_{q}\left\{
    [R'- I({q_{X}}, {q_{Y|X}}) ]^{+} 
    + D({q_{Y|X}} || W | {q_X})\right.
\notag \\
& \left. -\mu(\Gamma- {\rm E}_{q_X}[c(X)]) \right\}
 - \lambda(R'-R)
\notag\\
\leq & [R'-I({q_{X}^*}, {q_{Y|X}^*})]^{+} 
 +D({q^*_{Y|X}} || W | {q_X^*}) 
\notag\\
& -\mu(\Gamma- {\rm E}_{q^*_X}[c(X)]) - \lambda(R'-R)
\notag\\
 \stackrel{\rm (b)}{=} & 
D({q^*_{Y|X}} || W | {q_X^*}) + \lambda[R-I({q_{X}^*}, {q_{Y|X}^*})] 
\notag\\
&-\mu(\Gamma- {\rm E}_{q_X^*}[c(X)])
=G_{\rm DK}^{(\mu, \prmt)}(R,\Gamma \Vl W).
\notag
\end{align}
Step (a) follows from (\ref{ineq_Gsp}). 
Step (b) follows from the choice of $R'=I({q_{X}^*}, {q_{Y|X}^*})$.
$\quad$\hfill\IEEEQED

We can show that 
${G}_{\rm AR}(R,\Gamma\Vl W)$ 
and ${G}_{\rm AR}^{(\mu,\lambda)}(R,\Gamma\Vl W)$ 
satisfies the following property.
{\renewcommand{\prmt}{\lambda} 
\renewcommand{\barprmt}{1-\lambda} 
\begin{pr}{\label{pr:pr4z}}$\quad $
\begin{itemize}
\item[\rm{a)}] 
The function ${G}_{\rm AR}(R,\Gamma\Vl W)$ is monotone increasing 
function of $R$ and is positive if and only if $R>C(\Gamma| W)$. 
\item[\rm{b)}]
For $y\in{\cal Y}$, set
$$
\Lambda(y) \defeq \sum_{x\in{\cal X}}
q_{X}(x)\left\{W(y|x){\rm e}^{-\mu \lambda c(x)}\right\}^{\frac{1}{\barprmt}}\,. 
$$
Then, for $\lambda\in (0,1]$, necessary and sufficient conditions 
on the probability distribution $q_{X}\in {\cal P}({\cal X})$ that minimizes 
${J}^{(\mu,\prmt)}(q_{X}\Vl W)$ is 
$$
\sum_{y\in{\cal Y}}\left\{W(y|x)
{\rm e}^{-\mu \lambda c(x)}\right\}^{\frac{1}{\barprmt}}
     {\Lambda(y)}^{-\prmt}
\leq \sum_{ y \in {\cal Y} }\Lambda(y)^{\barprmt}
$$
for any $x\in {\cal X}$ with equality if $q_{X}(x)\neq 0$. 
\end{itemize}
\end{pr}
}

We now proceed to the proof of Proposition \ref{pro:pro1}. We first prove 
(\ref{eqn:mainEq00a}) and (\ref{eqn:mainEq00}) in Proposition \ref{pro:pro1}.

{\it Proofs of 
(\ref{eqn:mainEq00a}) and (\ref{eqn:mainEq00}) 
in Proposition \ref{pro:pro1}:} 
We first prove $G_{\rm DK}^{(\muprmt,\prmt)}(R,\Gamma|{\NoizeVar})$
$=G_{\rm AR}^{(\mu,\prmt)}(R,\Gamma|{\NoizeVar})$.
For a given joint distribution 
$$
(q_X,q_{Y|X})=
\left\{
q_{X}(x)q_{Y|X}(y|x) 
\right\}_{(x,y)\in {\cal X}\times {\cal Y}}\,,
$$
we introduce the stochastic matrix
$q_{X|Y}$ 
$=\left\{\empty{}q_{X|Y}(x|y)\right\}$ ${}_{(x,y) \in {\cal X} \times{\cal Y}}$ 
and the probability distribution 
$\empty{}q_{Y}$ 
$=\left\{\empty{}q_{Y}(y)\right\}$ ${}_{y \in {\cal Y}}$ 
by 
$$
q_{X}(x)q_{Y|X}(y|x) 
=\empty{}q_{Y}(y)\empty{}q_{X|Y}(x|y),\quad (x,y) 
\in {\cal X} \times{\cal Y}.
$$
The above $q_{X|Y}$ is called a backward channel. 
Using $(q_{Y},$ $q_{X|Y})$, we obtain the following 
chain of equalities: 
\beqa
& &-\prmt\left\{I(q_X,q_{Y|X})-\mu{\rm E}_{q_X}[c(X)]\right\}+D(q_{Y|X}||W|q_X)
\nonumber\\
&=&\prmt D(\empty{}q_{X|Y}||q_X |\empty{}q_{Y})
    +D(\empty{}q_{Y},\empty{}q_{X|Y}||q_X,W) 
\nonumber\\
& &+\muprmt {\rm E}_{(q_{Y},q_{X|Y})}[c(X)]
\nonumber\\
&=&\sum_{y\in{\cal Y}}\sum_{x\in{\cal X}} \empty{}q_{Y}(y)\empty{}q_{X|Y}(x|y)
\log\left\{\frac{\empty{}q_{X|Y}^{-\prmt}(x|y)}{q_X^{-\prmt}(x)} 
    \right\}
\nonumber\\
& & +\sum_{y\in{\cal Y}}\sum_{x\in{\cal X}}\empty{}q_{Y}(y)\empty{}q_{X|Y}(x|y)
\nonumber\\
& &\qquad\times
    \log \left\{ 
     \frac{\empty{}q_{X|Y}(x|y) \empty{}q_{Y}(y)}
     {q_X(x)W(y|x){\rm e}^{-\muprmt c(x)}} 
     \right\}
\nonumber\\
&=&\sum_{y\in{\cal Y}}\sum_{x\in{\cal X}}\empty{}q_{Y}(y)\empty{}q_{X|Y}(x|y)
\nonumber\\
& &\qquad\times
\log\left\{\frac{\empty{}q_{X|Y}^{\barprmt}(x|y)}
{q_X^{\barprmt}(x)W(y|x){\rm e}^{-\muprmt c(x)}} 
    \right\}
\nonumber\\
& &+\sum_{y\in{\cal Y}}\empty{}q_{Y}(y) \log \empty{}q_{Y}(y)
\nonumber\\
&=&\barprmt\sum_{y\in{\cal Y}}\sum_{x\in{\cal X}}
   \empty{}q_{Y}(y)\empty{}q_{X|Y}(x|y)
\nonumber\\
& &\qquad\times
    \log\left\{
        \frac{\empty{}q_{X|Y}(x|y)}
             {q_X(x)\left\{W(y|x){\rm e}^{-\muprmt c(x)}\right\}^{ 1+\lambda}} 
        \right\}
\nonumber\\
& &+\sum_{y\in{\cal Y}}\empty{}q_{Y}(y) \log \empty{}q_{Y}(y)
\nonumber\\
&=&\barprmt D(\empty{}q_{X|Y}||\hat{q}_{X|Y}|\empty{}q_{Y})
    +D(\empty{}q_{Y}||\hat{q}_{Y}) 
\nonumber\\
& &
- J^{(\mu,\prmt)}(q_X \Vl W). \label{eqn:equal0}
\eeqa
where 
$\hat{q}_{X|Y}$ 
$=\left\{\hat{q}_{X|Y}(x|y)\right\}$ 
${}_{(x,y) \in {\cal X} \times{\cal Y}}$ 
is a stochastic matrix whose components are 
\beqa
& &\hat{q}_{X|Y}(x|y)
\nonumber\\
&=&\frac{1}{\Lambda(y)}q_{X}(x){
\left\{W(y|x){\rm e}^{-\muprmt c(x)}\right\}
}^{1+\lambda}\,,
\:\:(x,y) \in {\cal X} \times{\cal Y}
\label{eqn:dist0}\qquad
\eeqa
and $\hat{q}_Y$ $=\left\{\hat{q}_Y(y)\right\}_{y\in{\cal Y}}$ 
is a probability distribution whose components are 
\beq
\hat{q}_Y(y)=\frac{\Lambda(y)^{\barprmt}}
         {\sum_{y\in{\cal Y}} \Lambda(y)^{\barprmt}}\,,
\quad y \in {\cal Y}.
\label{eqn:dist1}
\eeq
Hence, by (\ref{eqn:equal0}) and the non-negativity of divergence, 
we obtain 
$$
G_{\rm DK}^{(\muprmt,\prmt)}(R,\Gamma, q_X \Vl W)\geq 
{\empty}{G}_{\rm AR}^{(\mu,\prmt)}(R,\Gamma, q_X\Vl W)
$$ 
for any $q_X \in{\cal P}({\cal X})$. Next, 
we prove 
$$
G_{\rm DK}^{(\muprmt,\prmt)}(R,\Gamma \Vl W)= 
{\empty}{G}_{\rm AR}^{(\mu,\prmt)}(R,\Gamma\Vl W).
$$ 
To this end it suffices to show that for any $\lambda \geq 0$,
$$
{G}_{\rm DK}^{(\muprmt,\prmt)}(R,\Gamma \Vl W) 
\leq {\empty}{G}_{\rm AR}^{(\mu, \prmt)}( R,\Gamma \Vl W).
$$
Let $q_X$ be a probability distribution that attains the minimum
of ${\empty}{G}_{\rm AR}^{(\mu,\prmt)}(R,\Gamma, q_X \Vl W)$. 
Then, by Property \ref{pr:pr4z}, we have 
\beq
\sum_{y\in{\cal Y}}
\left\{W(y|x){\rm e}^{-\mu \prmt c(x)}\right\}^{1+\lambda}
    {\Lambda(y)}^{-\prmt}
\leq \sum_{ y \in {\cal Y} }\Lambda(y)^{\barprmt}
\label{eqn:cond0}
\eeq
for any $x\in {\cal X}$ with equality if $q_X(x)\neq 0$. For 
$x\in {\cal X}$ with $q_X(x)> 0$ and $y\in {\cal Y}$, define 
the matrix 
${V}$ 
$=\left\{ V(y|x)\right\}$ ${}_{(x,y) \in {\cal X} \times{\cal Y}}$ 
by
\beq
V(y|x)=\frac{\hat{q}_Y(y)\hat{q}_{X|Y}(x|y)}{q_X(x)}, \quad 
(x,y) \in {\cal X} \times{\cal Y}.
\label{eqn:vdist}
\eeq
By (\ref{eqn:dist0}) and (\ref{eqn:dist1}), each $V(y|x)$ has the 
following form:
\begin{align}
& V(y|x)=\frac{\Lambda(y)^{\barprmt}}
   {\sum_{y\in{\cal Y}} \Lambda(y)^{\barprmt}}
\notag\\
&\quad \times
   \frac{1}{\Lambda(y)}q_{X}(x)
\left\{W(y|x){\rm e}^{-\mu \prmt c(x)}\right\}^{1+\lambda}
   \cdot \frac{1}{q_X(x)}
\notag\\
&= \frac{
\left\{W(y|x){\rm e}^{-\muprmt c(x)}\right\}^{1+\lambda}
         \Lambda(y)^{-\prmt}}
        {\sum_{y\in{\cal Y}} \Lambda(y)^{\barprmt}}.
\label{eqn:dist4}
\end{align}
Taking summation of both sides of (\ref{eqn:dist4}) with respect 
to $y\in {\cal Y} $ and taking (\ref{eqn:cond0}) into
account, we obtain 
\beqno
\sum_{y\in{\cal Y}}{V}(y|x)
&=& \frac{\ds \sum_{y\in{\cal Y}}
\left\{W(y|x){\rm e}^{-\muprmt c(x)}\right\}^{1+\lambda}
\Lambda(y)^{-\prmt}}
     {\ds \sum_{y\in{\cal Y}} \Lambda(y)^{\barprmt}}
=1.  
\eeqno
The above equality implies that $V$ is a stochastic matrix. Furthermore, 
note that from (\ref{eqn:vdist}),
$$
q_{X}(x)V(y|x)=\hat{q}_{Y}(y)\hat{q}_{X|Y}(x|y),\quad (x,y)
\in {\cal X} \times{\cal Y}.
$$
Then, choosing $\empty{}q_{Y}=\hat{q}_Y,\empty{}q_{X|Y}=\hat{q}_{X|Y}$ 
in (\ref{eqn:equal0}), we have, for $\lambda \geq 0$, 
\beqno
& &{G}_{\rm DK}^{(\mu,\prmt)}(R, \Gamma \Vl W) 
\\
&\leq &\prmt\left\{(R-\mu \Gamma) -I(q_X,V)+ \mu{\rm E}_{q_X} [c(X)] \right\}
\\
& &+ D(V||W|q_X)
\nonumber\\ 
&=&\prmt(R-\mu \Gamma) - J^{(\mu,\lambda)}(q_X \Vl W)
={G}_{\rm AR}^{(\mu, \prmt)}(R,\Gamma \Vl W),
\eeqno
completing the proof. 
\hfill\IEEEQED

We prove (\ref{eqn:mainEq0}) in Proposition \ref{pro:pro1} 
by the two equalities (\ref{eqn:mainEq00a}) 
and (\ref{eqn:mainEq00}). 

{\it Proof of (\ref{eqn:mainEq0}) in Proposition \ref{pro:pro1}:} 
We first prove ${G}_{\rmOH}(R,\Gamma\Vl$ $ W)=$
${G}_{\rm DK}(R,\Gamma\Vl W)$. We have the following chain of inequalities:
\beqno
& &G_{\rmOH}(R,\Gamma \Vl W)
=\max_{\mu\geq 0,\lambda\geq 0}
G_{\rmOH}^{(\mu,\lambda)}(R,\Gamma \Vl W)
\\
&\MEq{a}&\max_{\scs \mu \geq 0}
    \max_{\scs
          \rho=\frac{\lambda}{1+\lambda}
         \scs \in [0,1)
   }
G_{\rm AR}^{(\mu,\rho)}(R,\Gamma \Vl W)=
G_{\rm AR}(R,\Gamma \Vl W).
\eeqno
Step (a) follows from (\ref{eqn:mainEq00a}) in Proposition \ref{pro:pro1}. 
We next prove 
${G}_{\rmOH}(R,\Gamma\Vl$ $W)=G_{\rm DK}(R,\Gamma|{\NoizeVar})$.
Let $q^*_X$ be an input distribution attaining $C(\Gamma|{\NoizeVar})$. 
Then, by the definition of $G_{\rm DK}^{(\mu,0)}(R,\Gamma|{\NoizeVar})$, 
we have
\beqa
G_{\rm DK}^{(\mu,0)}(R,\Gamma|{\NoizeVar}) 
&\leq&-\mu( \Gamma-{\rm E}_{q^*_X}[c(X)])\leq 0
\eeqa
for any $\mu\geq 0$. Hence we have 
$$
\max_{\mu\geq 0}G_{\rm DK}^{(\mu,0)}(R,\Gamma|{\NoizeVar})=0.
$$  
Then we have the following chain of inequalities:  
\beqno
& &G_{\rm DK}(R,\Gamma|{\NoizeVar})
=
\max_{\scs \mu\geq 0,\atop{\scs \lambda \in (0,1]}} 
G_{\rm DK}^{(\mu,\lambda)}(R,\Gamma|{\NoizeVar})
\\
&\MEq{a}&
\sup_{\scs \mu \geq 0,\atop{\scs \lambda \in (0,1)}} 
G_{\rm DK}^{(\mu,\lambda)}(R,\Gamma|{\NoizeVar})
=
\sup_{\scs \mu \geq 0,
\atop{\scs \lambda \in (0,1),
    \atop{\scs \beta= \frac{\lambda}{1-\lambda},\alpha=\frac{\mu}{\lambda}
         }
    }
} 
G_{\rm DK}^{(\mu,\lambda)}(R,\Gamma|{\NoizeVar})
\\
&=&
\sup_{\scs \mu=\frac{\alpha\beta}{1+\beta}\geq 0,
      \atop{\scs \lambda=\frac{\beta}{1+\beta}\in (0,1)}} 
G_{\rm DK}^{(\mu,\lambda)}(R,\Gamma|{\NoizeVar})
\\
&=&
\sup_{\scs \alpha \geq 0,\beta>0} 
G_{\rm DK}^{(\frac{\alpha\beta}{1+\beta},\frac{\beta}{1+\beta})}(R,\Gamma|{\NoizeVar})
\\
&\MEq{b}&
\sup_{\scs \alpha \geq 0,\beta > 0} G^{(\alpha,\beta)}(R,\Gamma|{\NoizeVar})
\MEq{c} \sup_{\scs \alpha \geq 0,\beta \geq 0} G^{(\alpha,\beta)}(R,\Gamma|{\NoizeVar})
\\
&=&G(R,\Gamma|{\NoizeVar}).
\eeqno
Step (a) follows from the continuous  
property of $G_{\rm DK}^{(\mu,\lambda)}($ $R,\Gamma|{\NoizeVar})$ for $\lambda=1$.
Step (b) follows from (\ref{eqn:mainEq00}) in Proposition \ref{pro:pro1}.
Step (c) follows from $G^{(\alpha,0)}(R,\Gamma|{\NoizeVar})=0$ for any $\alpha\geq 0$.
\hfill\IEEEQED
}

\newcommand{\ApdProOnePartB}{
\subsection{Proof of Property \ref{pr:prOne} part b)}


{\it Proof of Property \ref{pr:prOne} part b):} 
We first observe that
\beqa
&&{G}_{\rm DK}(R, \Gamma \Vl W)
\nonumber\\
&=&
\min_{\scs q: 
         {\rm E}_{q}[c(X)]\leq \Gamma
 }
\big\{ 
[R-I(q_X,q_{Y|X})]^+ +D(q_{Y|X} || W | q_X)\big\}
\nonumber\\ 
&=& 
\min_{\scs q: 
         {\rm E}_{q}[c(X)]\leq \Gamma
 }
\Theta(R,q \Vl W),
\eeqa
where we set
\beqno
&&\Theta(R,q \Vl W)
\\
&\defeq& [R-I(q_X,q_{Y|X})]^+ +D(q_{Y|X} || W | q_X) 
\\
&=&\max\{
\ba[t]{l}
R-I(q_X,q_{Y|X})+D(q_{Y|X} || W | q_X),\\ 
D(q_{Y|X}||W |q_X)\}. 
\ea
\eeqno
For each $i=0,1$, let $q^{(i)}_{XY}$ be a probability distribution 
that attains $G_{\rm DK}(R_i,\Gamma_i|{\NoizeVar})$.
By definition we have
\beq
G_{\rm DK}(R_i,\Gamma_i|{\NoizeVar})=\Theta(R_i,q^{(i)}|{\NoizeVar})
\mbox{ for }i=0,1.
\label{eqn:AszzzJ}
\eeq
For $\alpha_0\in [0,1]$, we set
$ 
q^{(\alpha)}_{XY}=\alpha_0 q^{(0)}_{XY}+\alpha_1 q^{(1)}_{XY},
$
where
$
\alpha_1=1-\alpha_0.
$
The quantities $q^{(\alpha)}_{X}$ and $q^{(\alpha)}_{Y|X}$ are probability 
and conditional probability distributions induced by 
$q^{(\alpha)}_{XY}$. 
Set $\Gamma_\alpha\defeq \alpha_0 \Gamma_0+\alpha_1 \Gamma_1.$
By the linearity of ${\rm E}_{q}[c(X)]$ with respect 
to $q$, we have that
\beqa
{\rm E}_{q^{(\alpha)}}[c(X)]
&=&\sum_{i=0,1}\alpha_i {\rm E}_{q^{(i)}}[c(X)]
\leq \Gamma_\alpha.
\label{eqn:AssCV2}
\eeqa
By the convex property of 
$-I(q_X,q_{Y|X})$ $+D(q_{Y|X} ||W|q_X)$ 
and $D(q_{Y|X} ||W|q_X)$ with respect to $q$, 
we have that
\beq
\left.
\ba{l}
\ds -I(q^{(\alpha)}_X,q^{(\alpha)}_{Y|X})
+D(q^{(\alpha)}_{Y|X} ||W|q^{(\alpha)}_X) 
\vspace*{1mm}\\
\quad \ds \leq
\sum_{i=0,1}\alpha_i\left[
-I(q^{(i)}_X,q^{(i)}_{Y|X})+D(q_{Y|X}^{(i)}||W|q^{(i)}_X) 
\right],
\\
\ds D(q^{(\alpha)}_{Y|X} ||W|q^{(\alpha)}_X)
\leq 
\sum_{i=0,1}\alpha_i D(q^{(i)}_{Y|X} ||W|q^{(i)}_X).
\ea
\right\}
\label{eqn:AssCV}
\eeq
Set $R_\alpha\defeq \alpha_0 R_0+\alpha_1 R_1$.
We have the following two chains of inequalities:   
\beqa
& &R_\alpha
 -I(q^{(\alpha)}_X,q^{(\alpha)}_{Y|X})
  +D(q^{(\alpha)}_{Y|X} ||W|q^{(\alpha)}_X) 
\nonumber\\
&\MLeq{a} & 
\sum_{i=0,1}\alpha_i\left[
 R_i-I(q^{(i)}_X,q^{(i)}_{Y|X})
+D(q^{(i)}_{Y|X} ||W|q^{(i)}_X)\right]
\nonumber\\
&\MLeq{b} & \sum_{i=0,1}\alpha_i\Theta(R_i,q^{(i)} \Vl W),
\label{eqn:aSddd}
\\
& &D(q^{(\alpha)}_{Y|X} ||W|q^{(\alpha)}_X) 
\MLeq{c} \sum_{i=0,1}\alpha_i D(q^{(i)}_{Y|X} ||W|q^{(i)}_X)
\nonumber\\
&\MLeq{d} & \sum_{i=0,1}\alpha_i \Theta(R_i,q^{(i)} \Vl W).
\label{eqn:aSddd2}
\eeqa
Steps (a) and (c) follow from (\ref{eqn:AssCV}). 
Steps (b) and (d) follow from the definition of 
$\Theta(R_i,q^{(i)} \Vl W),i=0,1.$
From (\ref{eqn:aSddd}) and (\ref{eqn:aSddd2}), 
we have that 
\beq
\Theta \left(\left. R_\alpha,q^{(\alpha)} \right| W \right)
\leq 
\sum_{i=0,1}\alpha_i\Theta(R_i,q^{(i)} \Vl W).
\label{eqn:aSddd3}
\eeq
Thus we have the following chain of inequalities
\beqno
& &G_{\rm DK}(R_\alpha,\Gamma_\alpha|{\NoizeVar})
=
\min_{\scs q:{\rm E}_{q}[c(X)] 
       \leq \Gamma_\alpha
 }
\Theta(R_\alpha,q \Vl W)
\\
&\MLeq{a}&
\Theta(R_\alpha,q^{(\alpha)}\Vl W)
\MLeq{b}\sum_{i=0,1}\alpha_i\Theta(R_i,q^{(i)} \Vl W)
\\
&\MEq{c}&\sum_{i=0,1}\alpha_iG_{\rm DK}(R_i,\Gamma_i|{\NoizeVar}).
\eeqno
Step (a) follows from (\ref{eqn:AssCV2}).
Step (b) follows from (\ref{eqn:aSddd3}). 
Step (c) follows from (\ref{eqn:AszzzJ}).
\hfill\IEEEQED 
}

\newcommand{\ApdProTwoZZPartB}{ 
\subsection{Proof of Property \ref{pr:prTwoZZ} part b)}


{\it Proof of Property \ref{pr:prTwoZZ} part b):} 
We set 
$$
\Theta^{(\mu)}(R,\Gamma, q \Vl W)
\defeq \Theta(R,q \Vl W)-\mu(\Gamma- {\rm E}_{q}[c(X)]).
$$
Then we have
$$
{G}_{\rm DK}^{(\mu)}(R, \Gamma \Vl W)
\nonumber\\ 
= \min_{\scs q}
\Theta^{(\mu)}(R,\Gamma, q \Vl W).
$$
For each $i=0,1$, let $q^{(i)}_{XY}$ be a probability distribution 
that attains $G_{\rm DK}(R_i,\Gamma_i|{\NoizeVar})$.
By definition we have
\beq
G_{\rm DK}(R_i,\Gamma_i|{\NoizeVar})=\Theta^{(\mu)}(R_i,\Gamma_i,q^{(i)}|{\NoizeVar})
\mbox{ for }i=0,1.
\label{eqn:bAszzzJ}
\eeq
For $\alpha_0\in [0,1]$, we set
$ 
q^{(\alpha)}_{XY}=\alpha_0 q^{(0)}_{XY}+\alpha_1 q^{(1)}_{XY},
$
where
$
\alpha_1=1-\alpha_0.
$
The quantities $q^{(\alpha)}_{X}$ and $q^{(\alpha)}_{Y|X}$ are probability 
and conditional probability distributions induced by 
$q^{(\alpha)}_{XY}$. 
By the convex property of 
\beqno
& &-I(q_X,q_{Y|X})+D(q_{Y|X} ||W|q_X) +\mu {\rm E}_q[c(X)]
\\
&&\mbox{and }D(q_{Y|X} ||W|q_X)+ \mu {\rm E}_q[c(X)]
\eeqno
with respect to $q$, we have that
\beq
\left.
\ba{l}
\ds -I(q^{(\alpha)}_X,q^{(\alpha)}_{Y|X})
+D(q^{(\alpha)}_{Y|X} ||W|q^{(\alpha)}_X) 
+\mu {\rm E}_{q^{(\alpha)}}[c(X)]
\vspace*{1mm}\\
\quad \ds \leq
\sum_{i=0,1}\alpha_i\left[
-I(q^{(i)}_X,q^{(i)}_{Y|X})+D(q_{Y|X}^{(i)}||W|q^{(i)}_X)\right. 
\vspace*{1mm}\\
\qquad \quad \qquad +\mu{\rm E}_{q^{(i)}}[c(X)]
\Big],
\\
\ds D(q^{(\alpha)}_{Y|X} ||W|q^{(\alpha)}_X)
+\mu{\rm E}_{q^{(\alpha)}}[c(X)]
\\
\quad \ds 
\leq 
\sum_{i=0,1}\alpha_i \left[
D(q^{(i)}_{Y|X} ||W|q^{(i)}_X)+\mu{\rm E}_{q^{(i)}}[c(X)]
\right].
\ea
\right\}
\label{eqn:bAssCV}
\eeq
Then we have the following two chains of inequalities:   
\beqa
& &
R_\alpha
 -I(q^{(\alpha)}_X,q^{(\alpha)}_{Y|X})
  +D(q^{(\alpha)}_{Y|X} ||W|q^{(\alpha)}_X) 
\nonumber\\
& &-\mu\left(\Gamma_\alpha - {\rm E}_{q^{(\alpha)}}[c(X)]
\right)
\nonumber\\
&\MLeq{a} & 
\sum_{i=0,1}\alpha_i\left[
 R_i-I(q^{(i)}_X,q^{(i)}_{Y|X})
+D(q^{(i)}_{Y|X}||W|q^{(i)}_X)\right.
\nonumber\\
& &-\mu(\Gamma_i-{\rm E}_{q^{(i)}}[c(X)])\Bigr]
\nonumber\\
&\MLeq{b} & \sum_{i=0,1}\alpha_i
\Theta^{(\mu)}(R_i,\Gamma_i,q \Vl W),
\label{eqn:baSddd}
\\
& &D(q^{(\alpha)}_{Y|X} ||W|q^{(\alpha)}_X) 
-\mu(\Gamma_\alpha- {\rm E}_{q^{(\alpha)}}[c(X)])
\nonumber\\
&\MLeq{c}& \sum_{i=0,1}\alpha_i 
\Bigl[
D(q^{(i)}_{Y|X} ||W|q^{(i)}_X)
-\mu( \Gamma_i - {\rm E}_{q^{(i)}}[c(X)]) \Bigr]
\nonumber\\
&\MLeq{d} & \sum_{i=0,1}\alpha_i
\Theta^{(\mu)}(R_i,\Gamma_i,q^{(i)} \Vl W).
\label{eqn:baSddd2}
\eeqa
Steps (a) and (c) follow from (\ref{eqn:bAssCV}). 
Steps (b) and (d) follow from the definition of 
$\Theta(R_i,\Gamma_i, q^{(i)} \Vl W),i=0,1.$
From (\ref{eqn:baSddd}) and (\ref{eqn:baSddd2}), 
we have that 
\beq
\Theta^{(\mu)}
\left(\left. R_\alpha, \Gamma_\alpha , q^{(\alpha)} \right| W \right)
\leq 
\sum_{i=0,1}\alpha_i
\Theta^{(\mu)}(R_i,\Gamma_i, q^{(i)} \Vl W).
\label{eqn:baSddd3}
\eeq
Thus we have the following chain of inequalities
\beqno
& &G_{\rm DK}^{(\mu)}(R_\alpha,\Gamma_\alpha|W)
=
\min_{\scs q
 }
\Theta^{(\mu)}(R_\alpha,\Gamma_\alpha,q \Vl W)
\\
&\leq&
\Theta^{(\mu)}(R_\alpha, \Gamma_\alpha, q^{(\alpha)} \Vl W)
\MLeq{a}\sum_{i=0,1}\alpha_i 
\Theta^{(\mu)}(R_i,\Gamma_i, q^{(i)} \Vl W)
\\
&\MEq{b}&\sum_{i=0,1}\alpha_i G_{\rm DK}^{(\mu)}(R_i,\Gamma_i|W).
\eeqno
Step (a) follows from (\ref{eqn:baSddd3}). 
Step (b) follows from (\ref{eqn:bAszzzJ}).
\hfill\IEEEQED 
}

\newcommand{\ApdProTwoBPartB}{
\subsection{Proof of Property \ref{pr:pr2} part b)}

{\it Proof of Property \ref{pr:pr2} part b): } 
Let ${\cal P}({\cal Y})$ 
be \irr{the} set of distribution function on 
${\cal Y}$
and 
${\cal P}({\cal Y|X})$ be 
\irr{the} set of conditional distribution on ${\cal X}$
given ${\cal Y}$. 
\irr{F}or fixed $q_Y \in {\cal P}({\cal Y})$, we set   
\begin{align*}
 &{G}_{\rm DK}^{(\mu)}(R,\Gamma, q_Y \Vl W) 
\notag\\
\defeq & 
\min_{
\scs q_{X|Y} \in {\cal P}({\cal X}|{\cal Y}):
}
\big\{ [R - I(q_Y, q_{X|Y})]^+ + D(q_{Y|X} ||W|q_X)
\\
& \hspace{1cm}-\mu(\Gamma-{\rm E}_{q_X}[c(X)])\big\}, 
\\
 &{G}_{\mSP}^{(\mu)}(R, \Gamma, q_Y \Vl W)
\notag\\
\defeq & 
\min_{
\scs q_{X|Y} \in {\cal P}({\cal X}|{\cal Y}):
            \atop{\scs I(q_Y, q_{X|Y}) \geq R
            }
}
\left\{D(q_{Y|X} ||W|q_X)
-\mu(\Gamma-{\rm E}_{q_X}[c(X)] )\right\}
\irr{,}
\\
&\hat{G}^{(\mu)}(R, \Gamma, q_Y \Vl W)
\notag\\
\defeq &
\min_{
\scs q_{X|Y} \in {\cal P}({\cal X}|{\cal Y}):
            \atop{\scs I(q_Y, q_{X|Y}) \leq R
            }
}
\big\{R - I(q_Y, q_{X|Y})+D(q_{Y|X} ||W|q_X)\\
&\hspace{1cm}-\mu(\Gamma-{\rm E}_{q_X}[c(X)] )
\big\}.
\end{align*}
It is obvious that 
\beqa
&& 
{G}_{\rm DK}^{(\mu)}(R,\Gamma, q_Y \Vl W) 
\nonumber\\
&&= 
\min\left\{ 
{G}_{\mSP}^{(\mu)}(R,\Gamma,q_Y \Vl W), \hat{G}(R,\Gamma,q_Y \Vl W)
\right\}\irr{,}
\label{eqn:aa0}\\
&& 
G_{\rm DK}^{(\mu)}(R,\Gamma|W) = \min_{q_{Y} \in {\cal P}({\cal Y})} 
{G}_{\rm DK}^{(\mu)}(R,\Gamma, q_Y \Vl W), \nonumber \\
&& {G}_{\mSP}^{(\mu)}(R,\Gamma, \Vl W) 
= \min_{q_Y \in {\cal P}({\cal Y})} 
  {G}_{\mSP}^{(\mu)}(R,\Gamma,q_Y \Vl W)\irr{.}
\label{eqn:aa1}
\eeqa
Since  
\begin{align*}
& -I(q_Y, q_{X|Y}) + D(q_{Y|X} || W | q_X)+\mu{\rm E}_q[c(X)] \\ 
%
& =  
\sum_{x \in \mathcal{X} }
q_{X|Y}(x|y)
\sum_{y \in \mathcal{Y} }
q_Y(y)
\left[\mu c(x) +
 \log \frac{ q_Y(y)}{W(y|x)}
\right]
\end{align*}
is a linear function of $q_{X|Y}$\irr{, 
it is maximized by} some 
$q_{X|Y}$ satisfying $I(q_Y, q_{X|Y}) = R$. 
Then, 
\beqno
& &\hat{G}^{(\mu)}(R, \Gamma,q_Y \Vl W)
\\
&=&
\min_{
\scs q_{X|Y} \in {\cal P}({\cal X}|{\cal Y}):
            \atop{\scs I(q_Y, q_{X|Y}) = R
           }
}
\left\{
D(q_{Y|X} ||W|q_X)-\mu(\Gamma-{\rm E}_{q}[c(X)])
\right\}.
\eeqno
Thus, by (\ref{eqn:aa0}), we have
$$
{G}_{\rm DK}^{(\mu)}(R, \Gamma, q_Y \Vl W) 
= {G}_{\mSP}^{(\mu)}(R,\Gamma, q_Y \Vl W).
$$
From the above equality and (\ref{eqn:aa1}), we obtain 
Property \ref{pr:pr2} part b). 
$\quad$\hfill\IEEEQED
}


\begin{thebibliography}{99}

\bibitem{Han98InfSpec}
T. S. Han, {\it Information-Spectrum Methods in Information
Theory. }Springer-Verlag, Berlin, New York, 2002. The Japanese 
edition was published by Baifukan-publisher, Tokyo, 1998.

\bibitem{ari}S. Arimoto,
``On the converse to the coding theorem for discrete memoryless 
channels,'' {\it IEEE Trans. Inform. Theory,} vol. IT-19, no. 3, pp. 
357-359, May 1973.

\bibitem{Dueck_Korner1979}
G.~Dueck and J.~K\"orner, ``Reliability function of a discrete memoryless
  channel at rates above capacity,'' \emph{IEEE Trans. Inform. Theory}, vol.
  IT-25, no.~1, pp. 82--85, 1979.

\bibitem{OohamaIeice15}Y. Oohama,
``On two strong converse theorems for discrete memoryless channels,'' 
\emph{IEICE Trans. Fundamentals}, vol. 98, no. 12, pp. 2471--2475, 2015.


\bibitem{Nagaoka01} 
H. Nagaoka, ``Strong converse theorems in quantum information theory,'' 
{\it Proceedings of ERATO Workshop on Quantum Information Science}, 
p. 33, 2001.

\bibitem{HayashiNagaoka03}M. Hayashi and H. Nagaoka, 
``General formulas for capacity of classical-quantum channels",
{\it IEEE Trans. Inform Theory}, vol. 49, no. 7, pp. 1753--1768, 2003.


\BiBArXiv
























\end{thebibliography}
\end{document}